\documentclass[aps,pra,superscriptaddress,showpacs]{revtex4}
\usepackage[intlimits]{amsmath}
\usepackage{amsfonts}
\usepackage{psfrag}
\usepackage{subfigure}
\usepackage[usenames]{color}
\usepackage{ifpdf}

\ifpdf
\usepackage{epstopdf}
\usepackage[pdftex]{hyperref}
\else
\usepackage[hypertex,ps2pdf,dvips,dvipdf,colorlinks,urlcolor=blue,citecolor=blue]{hyperref}
\fi

\advance\voffset by -1cm
\advance\hoffset by -0.5cm
\newcommand\be            {\begin{equation}}
\newcommand\bea           {\begin{equation}\begin{array}l\displaystyle}
\newcommand\ee            {\end{equation}}
\newcommand\bes           {\begin{subequations}}
\newcommand\esu           {\end{subequations}}
\newcommand\erf[1]        {\eqref{#1}}
\newcommand\labl[1]       {\label{#1}\ee}

\newcommand{\ud}{\mathrm d}

\newcommand{\bigx}[1]{\bBigg@{#1}}

\newcommand\eps           {\varepsilon}
\newcommand\fii           {\varphi}

\newcommand\mc            {\mathcal}

\newcommand\LL           {Lieb--Liniger }
\newcommand\p            {\partial}
\newcommand\psid         {\psi^{\dagger}}
\renewcommand\th         {\theta}
\newcommand\kb           {k_\text{B}}

\newcommand\vev[1]{{\langle#1\rangle}}
\newcommand\FF[1]{{\langle0\vert#1\vert\th_1,\dots,\th_n\rangle}}
\newcommand\no[1]{{\,:\!#1\!:\,}}

\def\3pt#1#2#3{{\langle{#1}\vert{#2}\vert{#3}\rangle}}

\newcommand\arxiv[2]      {\href{http://arXiv.org/abs/#1}{#2}}
\newcommand\doi[2]        {\href{http://dx.doi.org/#1}{#2}}
\newcommand\httpurl[2]    {\href{#1}{#2}}

\begin{document}

\title{1D Lieb--Liniger Bose Gas as Non-Relativistic Limit\\ 
of the Sinh--Gordon Model}

\author{M. Kormos}
\affiliation{SISSA and INFN, Sezione di Trieste, via Beirut 2/4, I-34151, 
Trieste, Italy}

\author{G. Mussardo}
\affiliation{SISSA and INFN, Sezione di Trieste, via Beirut 2/4, I-34151, 
Trieste, Italy}
\affiliation{International Centre for Theoretical Physics (ICTP), 
I-34151, Trieste, Italy}

\author{A. Trombettoni}
\affiliation{SISSA and INFN, Sezione di Trieste, via Beirut 2/4, I-34151, 
Trieste, Italy}

\pacs{67.85.−d, 05.30.Jp, 02.30.Ik, 03.75.Hh}

\begin{abstract}
\noindent
The repulsive Lieb--Liniger model can be obtained as the
non-relativistic limit of the Sinh--Gordon model: all physical
quantities of the latter model ($S$-matrix, Lagrangian and operators)
can be put in correspondence with those of the former.  We use this
mapping, together with the Thermodynamical Bethe Ansatz equations and
the exact form factors of the Sinh--Gordon model, to set up a compact
and general formalism for computing the expectation values of the
Lieb--Liniger model both at zero and finite temperatures. The
computation of one-point correlators is thoroughly detailed and when
possible compared with known results in the literature.
\end{abstract}
\maketitle

\section{Introduction}
\label{sec:intro}
\noindent
The physics of one-dimensional interacting bosons is well captured by
the Lieb--Liniger (LL) model \cite{LL}.  Despite its deceptive
simplicity,
this model has become a paradigmatic example of quantum integrable
systems since it proved to have a remarkable richness \cite{weiss}:
its Bethe Ansatz equations, for instance, can be explicitly derived and
used to study its equilibrium properties at zero and finite
temperatures \cite{yang}. The explicit analysis of the weak to strong
coupling crossover of this model has also set a
precise benchmark for approximate many-body techniques \cite{LL,weiss}. The
efforts done for the computation of the correlation functions of the
LL model have also greatly stimulated the development of new and
general formalisms, such as the quantum inverse scattering method
\cite{korepin} or the bosonization approach \cite{haldane}.

Nowadays a renewed interest in the LL model has been triggered by its
accurate experimental realization
\cite{paredes04,kinoshita04,vandruten08,nagerl}: in quasi
one-dimensional traps, the excitations in the transverse directions are
effectively frozen and, moreover, the coupling of the ultracold bosons
to the external environment can be made very weak
\cite{olshanii98}. These recent experimental advances have opened new
perspectives in the field of strongly correlated quantum systems: in
such a highly controllable set-up it is, in fact, possible to
thoroughly investigate problems of general nature concerning quantum
extended systems, such as the dynamics of integrable systems in the
presence of small non-integrable perturbations \cite{DMS} (e.g.,
three-body interactions and/or a weak external trapping potential),
the issue of thermalization in quantum integrable and non-integrable
systems \cite{olshanii2} and the behavior of various susceptibilities
and response functions.

The key quantities to answer all these questions are the correlation
functions of the LL model. Despite the integrability of the model,
their explicit computation turned out to be an interesting theoretical
challenge . For this reason many different approaches have been
developed over the years to tackle different aspects of this difficult
problem: a partial list includes bosonization (which gives the correct
long-distance behavior of correlation functions)
\cite{haldane,cazalilla,giamarchi06}, quantum Monte Carlo simulations
\cite{giorgini}, algebraic Bethe Ansatz \cite{slavnov},
analytical-numerical methods based on the exact Bethe Ansatz solution
\cite{caux,guan}, Bogoliubov weak- \cite{castin} and strong-coupling
methods \cite{gangardt}, renormalization group \cite{dunjko09},
numerical results using stochastic wave-functions \cite{carusotto} and
imaginary time simulations \cite{deuar,deaur2}. Exact results based on
the Yang--Yang equations and the Hellmann--Feynman theorem are
presently available for local two-body correlations
\cite{gangardt,kheruntsyan}, while the local three-body correlations
were determined at zero temperature in \cite{cheianov}.
 
A new method was proposed recently to compute expectation values in
the LL model \cite{KMT}: it exploits a different route from all the
previous approaches for it is based on an exact mapping between the
non-relativistic LL model and the relativistic integrable Sinh--Gordon
(sh-G) model. This proposal not only provides a remarkable
simplification of the problem but applies equally well both at zero
and finite temperatures. In a nutshell, the logical steps on which the
method is based are the following:
\begin{enumerate}
\item Due to the relativistic invariance and quantum integrability 
of the Sh-G model, it is possible to set up functional
equations \cite{smirnov,watson} for the matrix elements of its local
operators on the asymptotic states -- known as {\em form factors} --
and to find their exact solutions \cite{FMS,mussardo}.
\item The finite temperature and finite density effects of the Sh-G model
can also be controlled by solving the Thermodynamical Bethe Ansatz
(TBA) equations \cite{zam,klass}.
\item In a proper non-relativistic limit of the sh-G model, all quantities
of this theory -- $S$-matrix, Lagrangian, form factors, TBA equations
and so on -- reduce to those of the LL model and therefore can be used to
establish an explicit mapping between the two models. In particular,
from the exact and known expressions of the form factors of the sh-G
model we can explicitly obtain the matrix elements of the
operators of the LL model we are interested in.
\item To actually compute the LL correlation functions we have to take into account 
another aspect of the problem, that is, that the LL correlation functions
refer to the ground state of the gas at a finite density and at a finite temperature, 
while those of the sh-G model refer to the vacuum state (i.e. the state without any
particles). This apparent difficulty can be, however, readily overcome
by using the LeClair--Mussardo formalism \cite {leclair} that, as a
matter of fact, is based on the same quantities mentioned above,
i.e., the form factors and the TBA equations.
\end{enumerate}

In this paper we provide the details of the mapping between the sh-G 
and the LL models first established in \cite{KMT}, presenting 
additional results and focusing 
values. The paper is organized as follows: in Section \ref{sec:models} we
introduce the LL and sh-G models and we recall their TBA equations. In
Section III we discuss the form factors of the sh-G model and their
correspondence with the operators of the theory. In
Section \ref{sec:limit} we explore in detail the non-relativistic regime
of the sh-G model and we explain how the LL model emerges in this
limit: the mapping between the two models is discussed at the level of
their $S$-matrices, Lagrangians and TBA equations. In
Section \ref{sec:LL1pt} we describe the method of calculating local
correlation functions in the LL model, providing a detailed derivation
of the main result, formula \erf{eq:formula}.  Then we present the
explicit computation of LL one-point correlators at zero and finite
temperatures, together with their comparison with known exact and
approximate results available in literature. Our conclusion and
outlooks are given in Section \ref{sec:concl}. Supplementary material is
presented in the Appendices, where we also list the non-relativistic
limit of the sh-G form factors for the first few cases.

\section{The Models}
\label{sec:models}
\noindent 
In this section we recall the main properties of the LL and sh-G
models and we also discuss the expressions of their free energy and
$T=0$ ground-state energy obtained by the TBA equations.

\subsection{Lieb--Liniger Hamiltonian}
\noindent
The LL Hamiltonian describes $N$ non-relativistic bosons of mass $m$
in one dimension, interacting via a two-body repulsive
$\delta$-potential:
\be
\label{HAM_LL}
H=-\frac{\hbar^2}{2m}\sum_{i=1}^N\frac{\p^2}{\p x_i^2}+2\lambda\,
\sum_{i<j}\delta(x_i-x_j)\,. 
\ee
For cold atomic gases the quantity $\lambda>0$ in the Hamiltonian (\ref{HAM_LL}) can be
determined in terms of the parameters of the three-dimensional Bose
gas in the quasi-one-dimensional limit \cite{olshanii98}. The
effective coupling constant of the LL model is given by the
dimensionless quantity
\begin{equation}
\gamma=\frac{2m\lambda}{\hbar^2n}\,, 
\label{eq:gamma}
\end{equation}
where $n=N/L$ is the density of the gas ($L$ is the length of the
system). The limit $\gamma \ll 1$ is the weak coupling limit and in
this regime it is known that the Bogoliubov approximation -- obtained
by linearizing the Gross--Pitaevskii equation -- gives a good estimate
of the ground-state energy of the system \cite{LL}. For large $\gamma$
one approaches the Tonks--Girardeau limit \cite{girardeau60} and
recently it has become an interesting question, also from the
experimental point of view, to study the crossover between the two
regimes \cite{weiss}. In the LL model temperatures are usually
expressed in units of the quantum degeneracy temperature
\be
\kb T_\text{D}=\frac{\hbar^2 n^2}{2m}\,,
\labl{eq:tau} 
in the following we use the scaled temperature $\tau=T/T_\text{D}$.

In second quantized formalism the non-relativistic field theory describing 
bosons interacting via a $\delta$-potential is defined by the Hamiltonian 
\cite{korepin}
\be
{\cal H}\,=\,\int\mathrm{d}x\,\left(\frac{\hbar^2}{2m}\frac{\p\psid}{\p x}
\frac{\p\psi}{\p x}+\lambda\,\psid\psid\psi\psi\right)\,,
\labl{eq:HLL}
where the complex Bose field $\psi(x,t)$ satisfies the 
canonical commutation relations
\be 
[\psi(x,t),\psid(x',t)]=\delta(x-x') \; ,
\,\,\,\,\,\, 
[\psi(x,t),\psi(x',t)]=0 \;. 
\label{CRPSI}
\ee
The Lagrangian density associated to the field theory Hamiltonian
${\cal H}$ is 
\be
{\cal L}= i\,\frac\hbar2\left(\psid\frac{\p\psi}{\p t}
- \frac{\p\psid}{\p
  t}\psi\right) -
\frac{\hbar^2}{2m} \frac{\partial\psid}{\partial x} 
\frac{\partial\psi}{\partial x} - \lambda\, \psid\psid\psi\psi\;.
\label{LagrangianNLS}
\ee
Restricting to the subspace of the Hilbert space where the number of
particles $N$ is fixed, the equation of motion for the field $\psi$
translates into the eigenvalue equation of the LL many-body
Hamiltonian \eqref{HAM_LL}:
\be 
H \,\chi_{N}(x_1,\dots,x_N)\,=
\,E_N \,\chi_N(x_1,\dots,x_N)\,.
\label{eigenLLH}
\ee
As shown by Lieb and Liniger in their original paper \cite{LL}, the
eigenvalue problem (\ref{eigenLLH}) can be solved in terms of a Bethe
Ansatz. Let us recall the main steps because they will lead
us to the definition of the $S$-matrix of the LL model, a basic
quantity of its dynamics and of our following discussion. 
 One can easily see that Eq.~(\ref{eigenLLH})
is just a free Schr\"odinger equation in the domain where the
coordinates of the particles are all distinct. However, we have to
enforce the usual boundary conditions for a $\delta$-function potential,
that is, the discontinuity of the gradient of the wave function when two
coordinates coincide
\be
\frac{\hbar^2}{2m}\left(\frac{\p}{\p x_j}-\frac{\p}{\p
  x_k}\right)\chi_N|_{x_j=x_k+\epsilon}=\lambda\,\chi_N|_{x_j=x_k}\,.
\label{cuspcond}
\ee
If we denote by $R_1$ the subset of the configuration space where
$x_1<x_2<\dots<x_N$ , the solution of the equations in $R_1$
is given by the Bethe wave function 
\be
\chi_N(x_1,x_2,\dots,x_N)=\sum_P a(P)\,e^{\frac{i}\hbar\sum_{j=1}^N P(k_j)x_j}
\,,
\ee
where $\sum_P$ denotes a sum over permutations of the momenta $\{k_1,\dots,k_n\}$ that characterize the state. For configurations
outside $R_1$ the solution is easily obtained using the symmetry of
$\chi_N$ with respect to the $x_i$. The coefficients in the sum are related by the boundary conditions 
(\ref{cuspcond}). For the permutations  
$P:(k,l,k_{\alpha_3},\dots,k_{\alpha_N})$ and
$Q:(l,k,k_{\alpha_3},\dots,k_{\alpha_N})$ the relation between the
corresponding coefficients is
\be
a(Q)=\frac{k-l-i\frac{2m}{\hbar}\,\lambda}{k-l+i\frac{2m}{\hbar}\,\lambda}\,a(P)\,.
\ee
Hence the wave function gets multiplied by the factor $a(Q)/a(P)$
whenever two particles with momenta $p_1= k$ and $p_2= l$
are exchanged. This exchange is equivalent to a scattering process of
the two particles and therefore 
the two-body $S$-matrix of the \LL model is expressed by 
\be
S_{\text{LL}}(p,\lambda)=\frac{p-i\frac{2m}{\hbar}\,\lambda}{p+i\frac{2m}{\hbar}\,\lambda}\,,
\label{SmatrixLLM}
\ee
where $p=p_1-p_2$ is the momentum difference. 

Given the integrability of the model, all its physical properties can be essentially derived from its two-body $S$-matrix (\ref{SmatrixLLM}). This quantity, for instance, determines the thermodynamics of the model, as shown originally by Yang and Yang \cite{yang}. In the limit $N \rightarrow \infty$, $L \rightarrow \infty$
with the density $n$ fixed, the discrete energy levels of the system
get encoded in an energy level density function $\tilde\rho(p)$ and in
the density $\tilde\rho^\text{(r)}(p)$ of the occupied levels. Notice
that we are going to use a tilde $\tilde{}\,$ for 
the quantities in the non-relativistic LL TBA, while the corresponding 
quantities in the TBA for the sh-G model will be later denoted without this tilde. 
The ratio between the two densities $\tilde\rho$ and $\tilde\rho^\text{(r)}$ defines the pseudo-energy $\tilde\eps(p)$ through the relation
\be
\frac{\tilde\rho(p)}{\tilde\rho^\text{(r)}(p)}\,=\,1+e^{\tilde\eps(p)}\,,
\label{pseudo_energy}
\ee
and this quantity, together with the densities, satisfies the coupled set of integral equations
\bes
\begin{align}
2\pi\tilde\rho(p)&=\frac1\hbar + \int_{-\infty}^\infty\ud
p'\,\tilde\fii(p-p')
\,\tilde\rho^{\text{(r)}}(p')\,,\\
\tilde\eps(p)& = -\frac{\tilde\mu}{\kb T}+\frac{p^2}{2m\kb T}-
\int_{-\infty}^\infty\frac{\ud p'}{2\pi}\, 
\tilde\fii(p-p')\log\left(1+e^{-\tilde\eps(p')}\right)\,,\\
n&=\int_{-\infty}^\infty \tilde\rho^\text{(r)}(p)\,\ud p\,,
\end{align}
\label{eq:YYTBA1}
\esu
~\hspace{-3mm}where $\tilde\mu$ is the chemical potential, $T$ is the
temperature and  
$\kb$ is the Boltzmann constant. The kernel $\tilde\fii(p)$ that drives
all the integral equations entirely follows from the $S$-matrix (\ref{SmatrixLLM})
\be
\tilde\fii(p)\,=\, -i \frac{\partial}{\partial p} \log S_{LL}(p)\,=\,\frac{4\hbar m\lambda}{\hbar^2p^2+4m^2\lambda^2}\,\,\,. 
\labl{eq:fiiLL}
Once the TBA integral equations (\ref{eq:YYTBA1}) are
solved, the ground state energy $\tilde E$ and the free energy $\tilde
F$ of the system are
expressed as
\bes
\begin{align}
\frac{\tilde E}L& = 
\int_{-\infty}^\infty\ud p\,\frac{p^2}{2m}\,\tilde\rho^\text{(r)}(p)\,,\\
\frac{\tilde F}L&=
\tilde\mu n-\frac{\kb T}{2\pi\hbar}\int_{-\infty}^\infty\ud p\, 
\log\left(1+e^{-\tilde\eps(p)}\right)\,.
\end{align}
\label{eq:YYEs1}
\esu

At zero temperature the energy level density 
gets a compact support, i.e. it is different from zero only on an 
interval (which we denote by $[-B,B]$) and, correspondingly, the TBA equations simplify as  
\bes
\begin{align}
2\pi\tilde\rho^{\text{(r)}}(p)&=\frac1\hbar + \int_{-B}^B\ud
p'\,\tilde\fii(p-p') 
\,\tilde\rho^{\text{(r)}}(p')\,\,\,,\\
\tilde\eps_0(p)& = -\tilde\mu+\frac{p^2}{2m}+
\int_{-B}^B\frac{\ud p'}{2\pi}\, 
\tilde\fii(p-p')\,\tilde\eps_0(p')\,\,\,,
\end{align}
\label{eq:YYT01}
\esu
where $\tilde\eps_0(p)=\lim_{T\to0}\kb T\,\tilde\eps(p)$ and the boundary value $B$ is determined by
the normalization condition
\be
n\,=\int_{-B}^{B} \tilde\rho^\text{(r)}(p)\,\ud p\,.
\ee 

The TBA equations of the sh-G model will be described in the next
section where the physical meaning of the pseudo-energy will also be
discussed. In Section \ref{sec:TBAlim} we show that the TBA equations
and the excitation spectrum of the LL model can be obtained in a
proper limit from those of the sh-G model.

\subsection{The Sinh--Gordon model}
\label{sec:sg}
\noindent
The sh-G model is an integrable relativistically invariant field
theory in $1+1$ dimensions defined by the Lagrangian density
\be
\mc{L}= \frac12\left(\frac{\p\phi}{c\,\p
  t}\right)^2-\frac12\left(\frac{\p\phi}{\p x}\right)^2 -
\frac{m_0^2c^2}{g^2\hbar^2}\left(\cosh(g\phi)-1\right)\,, 
\ee
where $\phi=\phi(x,t)$ is a {\em real} scalar field, $m_0$ is a mass
scale and $c$ is the speed of light. The parameter $m_0$ is related to
the physical (renormalized) mass $M$ of the particle by \cite{babkar}
\be
m_0^2\,=\,M^2\frac{\pi\alpha}{\sin(\pi\alpha)}\,.
\labl{eq:mu_m}
The explicit presence of the speed of light $c$ will help us in
studying later the non-relativistic limit of this theory (see Section IV). Despite the relativistic nature of the sh-G model, its integrability (supported by the
existence of an infinite number of conservation laws) implies the absence of 
particle production processes and that its $n$-particle scattering amplitudes are purely elastic. Moreover, they factorize into $n (n-1)/2$ two-body $S$-matrices. 
The energy $E$ and the momentum $P$ of a particle can be written as $E=M c^2 \cosh\th$, $P=M
c \sinh\th$, where $\th$ is the rapidity. In terms of the particle
rapidities, the two-body $S$-matrix is given by \cite{ari}:
\be
S_{\text{sh-G}}(\th,\alpha)=\frac{\sinh\th-i\,\sin(\alpha\pi)}{\sinh\th+i\,\sin(\alpha\pi)}\,,
\ee
where $\th$ is the rapidity difference and $\alpha$ is the dimensionless renormalized coupling constant
\be
\alpha\,=\,\frac{\hbar c\,g^2}{8\pi+\hbar c\,g^2}\,.
\labl{eq:alpha}

As in the LL model, the two-body $S$-matrix of the sh-G model fully
encodes its physical properties, in particular its thermodynamics. Its
derivation is quite similar to the one of the LL model, the only
difference being the relativistic kinematics \cite{zam, klass}.  Let us
briefly discuss the TBA equations of the sh-G model both at finite
temperature and at finite particle density $n$. Note initially that,
although the sh-G model is a relativistic theory, its quantum
integrability implies the conservation of the number of particles and
therefore it makes sense to associate a wave-function to the
$N$-particle state. Therefore, the starting point of the TBA approach
is the quantization of the rapidities of the $N$-particle state on an interval $L$ with periodic boundary conditions, given by 
\be
M c L \sinh\th_i + \hbar\sum_{j\neq i}^N \chi(\th_i-\th_j)\,=\, 2\pi {\mc N}_i\hbar\,,
\ee
where the ${\mc N}_i$ are (positive or negative) integers and 
$\chi(\th)=-i\log S_{\text{sh-G}}(\th)$ is the phase shift of the
two-body scattering process of  the sh-G model. 
In the thermodynamic limit ($N\to\infty$, $L\to\infty$,
$N/L=n=\text{fixed}$) we introduce for the left hand side of Eqs. (22)
the quantity
\be
J(\th)= Mc\sinh\th +
2\pi\hbar\,\int_{-\infty}^\infty\frac{\mathrm{d}\th'}{2\pi}\,
\chi(\th-\th')\rho^{\text{(r)}}(\th')\,\,\,,
\ee
and by differentiating it equations (22) turn into the integral equation
\be
\rho(\th)= \frac{Mc}{2\pi\hbar}\cosh \th +
\int_{-\infty}^\infty\frac{\mathrm{d}\th'}{2\pi}\,
\fii(\th-\th')\rho^{\text{(r)}}(\th')\,.
\labl{eq:rhos}
In the equations above
\be
\rho(\th)=\frac1{2\pi\hbar}\frac\p{\p\th}J(\th)
\ee
is the density of states,
$\rho^\text{(r)}(\th)$ is the density of the occupied states (both per unit length) and 
$\fii(\th)$ is the derivative of the phase shift
\be
\fii(\th)=\frac{\p\chi(\th)}{\p\th} =
-i\frac\p{\p\th}\log S_{\text{sh-G}}(\th)\,.
\ee
The pseudo-energy $\eps(\th)$ is introduced as in (\ref{pseudo_energy})
\be
\frac{\rho(\th)}{\rho^\text{(r)}(\th)}=1+e^{\eps(\th)}\,.
\labl{eq:eps} 
By minimizing the free energy 
\begin{multline}
F[\rho,\rho^\text{(r)}]  = E - T S = \\
L\int M \cosh\th \,\rho^\text{(r)}(\th) \, d\th  
 -  LT \int \left[\rho\,\log\rho - \rho^\text{(r)}\,\log  
\rho^\text{(r)} - (\rho - \rho^\text{(r)}) \, \log(\rho - \rho^\text{(r)}) \right]
\, d\th
\end{multline}
with respect to the densities $\rho(\th)$ and $\rho^\text{(r)}(\th)$, with the 
constraint (\ref{eq:rhos}), one arrives at the TBA equation for the pseudo-energy
\be
\eps(\th)= \frac{Mc^2}{\kb T}\cosh \th - \frac\mu{\kb T}-
\int_{-\infty}^\infty\frac{\mathrm{d}\th'}{2\pi}\,\fii(\th-\th')
\log\left(1+e^{-\eps(\th')}\right)\,.
\labl{eq:TBA}
Once this integral equation for $\eps(\th)$ has been solved, the
densities $\rho(\th)$ and $\rho^\text{(r)}(\th)$ are extracted from
equations \erf{eq:rhos} and \erf{eq:eps} while the chemical
potential is fixed by the constraint
\be
n=\int_{-\infty}^\infty\mathrm{d}\th\, \rho^{\text{(r)}}(\th)\,.
\labl{eq:n}
Any other thermodynamic quantity can be calculated from the free
energy (per unit length) of the system determined from the above
minimum principle and finally expressed by the formula
\be
f=\frac{F}L=
-\frac{\kb T}{2\pi\hbar}\int_{-\infty}^\infty\mathrm{d}\th\,Mc\cosh \th\,
\log\left(1+e^{-\eps(\th)}\right)+\mu n\,.
\labl{eq:f}

Let us consider now the $T\to0$ limit of the TBA equations. 
Due to the non-zero chemical potential,  we can assume 
that the function $\eps(\th)$
changes sign at the rapidity values $-\th^*$ and $\th^*$ so that it is
negative on the interval $I \equiv (-\th ^*,\th^*)$, zero at the limiting
points and positive everywhere else. For $T\to 0$, $\eps(\th)$ becomes
largely negative on the interval $I$ and largely positive outside. Then from
Eq.~\erf{eq:eps} we see that on the interval $(-\th ^*,\th^*)$ we have $\rho^\text{(r)}=\rho$ 
(and a filling fraction equal to $1$), while outside $\rho^\text{(r)}=0$ (and 
a filling fraction equal to $0$). The boundary value $\th^*$ is determined 
by the condition
\be
n=\int_{-\th^*}^{\;\th^*}\rho^\text{(r)}(p)\,\ud p\,.
\ee 
For $-\th^*<\th<\th^*$ equations \erf{eq:rhos} and
\erf{eq:TBA} become 
\bes
\begin{align}
\rho^\text{(r)}(\th)&= \frac{Mc}{2\pi\hbar}\cosh \th +
\int_{-\th^*}^{\th^*}\frac{\mathrm{d}\th'}{2\pi}\,
\fii(\th-\th')\,\rho^{\text{(r)}}(\th')\,,\\
\eps_0(\th)&= Mc^2\cosh \th-\mu+
\int_{-\th^*}^{\th^*}\frac{\mathrm{d}\th'}{2\pi}\,\fii(\th-\th')
\,\eps_0(\th')\,,\label{eq:epsT0}
\end{align}
\label{eq:TBAT0}
\esu
where $\eps_0(\th)=\lim_{T\to0}\kb T \eps(\th)$. 

Let us pause here to comment on the physical meaning of
$\eps(\th)$. If we raise one of the numbers ${\mc N}_i$ in the Bethe equation
that corresponds to a rapidity $\th_i$, to a larger value ${\mc N}'_i$,
which will correspond to a rapidity $\th'$, then the change in the
energy of the system is
\be 
\Delta E=\kb T(\eps(\th')-\eps(\th))\,. 
\ee 
This shows that $\eps(\th)$ describes the energy of the excitations
over the ground state, and it gives the dressed energy of the quasi-particles (hence the name pseudo-energy). If we demand that the ratio of the densities take the
usual form 
\be 
\frac{\rho^\text{(r)}(\th)}{\rho(\th)}=\frac1{e^{(E(\th)-\mu)/\kb T}+1} 
\ee 
then the excitation energy is fixed to be 
\be 
E(\th)=\kb T\,\eps(\th)+\mu\,.
\ee

\section{Form factor expansion for one-point correlators}
\label{sec:FF}
\noindent
Expectation values in an integrable relativistic field theory are
conveniently expressed in terms of the so-called {\em form
factors}.

The form factors we are going to use in this article are the ones
related to a quantum field theory, which in principle are different
from the quantities that share the same name in the Bethe ansatz
solution of integrable models. In the latter context they are defined
as matrix elements of operators between exact Bethe ansatz states,
while here we use instead the basis in the Hilbert space consisting of
multiparticle scattering states. However, these quantities turn out to
be closely related \cite{KMP}. In this section, for the sake of
completeness, we first review the definition and the main properties
of the form factors in a general relativistic field theory. Since our
final goal is to treat the sh-G model which is a theory with a single
type of gapped excitations and multi-particle states
without any bound states, we focus our attention on theories
of this type. (A well-known example of a theory with different types
of excitations and bound states is the Sine--Gordon
model \cite{zamzam}.) Later we present the explicit expressions of the
form factors of the sh-G model.

\subsection{Basic properties of form factors}
\begin{figure}[t]
\centerline{\scalebox{0.3}{\includegraphics{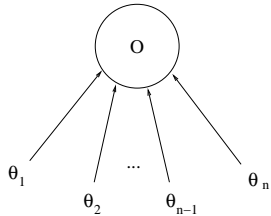}}}
\caption{Form factor of the operator ${\mathcal O}$.}
\label{fig:formfactor}
\end{figure}

\noindent
Consider a local operator $\mc O(x,t)$. Using the translation operator 
$U= e^{-i p_{\mu} x^{\mu}}$, where $t = x^{0}$ and $x=x^{1}$, we can always 
shift this operator to the origin ${\mc O(x,t)} =  
U^{\dagger} {\mc O(0,0)} U$. 
The matrix elements of $\mc O(0,0)$ between the vacuum and
a set of $n$-particle asymptotic states are called the $n$-particle
form factors of this operator (see Fig.~\ref{fig:formfactor})
\be
F_n^{\mc O}(\th_1,\th_2,\dots,\th_n)=\FF{\mc O(0,0)}\,.
\ee

For an operator of spin $s$, relativistic invariance implies that under a simultaneous 
shift in the rapidities its form factors behave as 
\be
F_n^{\mc O}(\th_1+\Lambda,\dots,\th_n+\Lambda)\,=\, e^{s\Lambda} F_n^{\mc O} (\th_1,\dots,\th_n)\,.
\labl{eq:asympt}
This equation indicates that the form factors of a scalar operator depend only on
the differences of rapidities, $\th_{ij}=\th_i-\th_j$. A generic matrix element of the operator ${\mc O(0,0)}$ can be expressed in terms of its 
form factors by using the translation operator and the crossing symmetry, 
which is implemented by an analytic continuation in the rapidity
variables
\be
\3pt{\th_1,\dots,\th_n}{\mc O(0,0)}{\beta_1,\dots,\beta_m}=F_{n+m}^{\mc O}(\beta_1,\dots,\beta_m,\th_1-i\pi,\dots,\th_n-i\pi)\,.
\ee
(If $\beta_i=\th_j$ for some $i$ and $j$, this formula gets
modified by contact terms.)
Hence, the knowledge of all form factors of an operator is equivalent to 
the knowledge of the operator itself, (i.e., how it acts on any state of the theory).

\begin{figure}[t]
\centerline{\scalebox{0.3}{\includegraphics{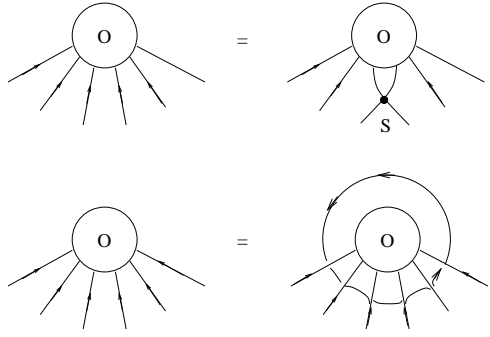}}}
\caption{Graphical form of the Watson equations.}
\label{fig:graficaFF}
\end{figure}

The form factors satisfy a set of functional and recursive equations,
which for integrable models makes it possible to find in many cases
their explicit expressions (for a review, see
\cite{smirnov,GMbook}). For a scalar operator the functional
equations (known as Watson equations \cite{watson}) come from unitarity and crossing
symmetry and their explicit expressions are
\bes
\begin{align}
F_n(\th_1,\dots,\th_i,\th_{i+1},\dots,\th_n)&=S(\th_i-\th_{i+1})\,F_n(\th_1,\dots,\th_{i+1},\th_i,\dots,\th_n)\,,\label{watson1} \\
F_n(\th_1+2\pi i,\dots,\th_n)&=\prod_{i=2}^nS(\th_i-\th_1)\,F_n(\th_1,\dots,\th_n)\,\,\,.
\label{watson2}
\end{align}
\label{eq:watson}
\esu
Their graphical representations are given in Fig.~\ref{fig:graficaFF}. 

\begin{figure}[b]
\centerline{\scalebox{0.3}{\includegraphics{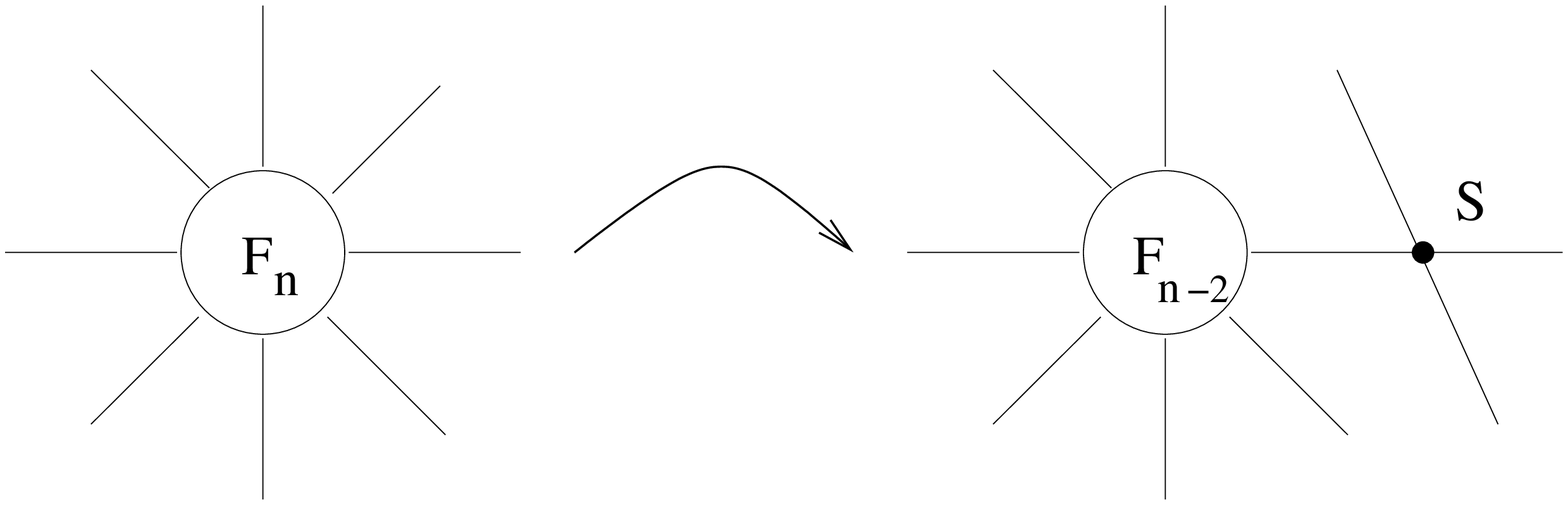}}}
\caption{Recursive equation of the kinematical poles.}
\label{fig:graficareckin}
\end{figure}

The recursive equations, on the other hand, come from the pole
structure of one-particle intermediate states. The form factors of integrable theories 
have, in general,
two kinds of simple poles in the strip $0<\mathrm{Im}\,\th_{ij}<2\pi$
and except for these singularities they are analytic in the
strip. The first kind of poles corresponds to kinematical
singularities at $\th_{ij}=i\pi$ and their residues give rise to a set
of recursive equations between the $n$-particle and the $n+2$-particle
form factors (see Fig.~\ref{fig:graficareckin})
\be
-i\mathop{\textrm{Res}}_{\tilde\th=\th}
F_{n+2}(\tilde\th+i\pi,\th,\th_1,\dots,\th_n)= \left(1-\prod_{i=1}^nS(\th-\th_i)\right)F_n(\th_1,\dots,\th_n)\,.
\label{reskin}
\ee
The second kind of poles is instead related to the bound states of the theory.  
Since there are no bound states in the sh-G model, there are no such poles 
in the form factors of this theory and we do not need to write here 
the corresponding residue equations. 

The general solution of the Watson equations \erf{eq:watson} can be written as
\be
F_n(\th_1,\dots,\th_n)=K_n(\th_1,\dots,\th_n)\prod^n_{i<j}F_\text{min}(\th_{ij})\,,
\label{generalsolutionW}
\ee
where 
the factors $K_n(\th_1,\dots,\th_n)$ are completely symmetric and
$2\pi i$-periodic functions in all
$\th_i$, and
$F_\text{min}(\th)$ is an analytic function in $0\le\mathrm{Im}\,\th\le\pi$ 
(without zeros and poles in this strip) which tends to a constant value for large values 
of $\th$. This function satisfies the equations \bes
\begin{align}
F_\text{min}(\th)&=S(\th)\,F_\text{min}(-\th)\,,\\
F_\text{min}(i\pi-\th)&=F_\text{min}(i\pi+\th)\,,  
\end{align}
\label{eq:Fmineq}
\esu
\hspace{-1.8mm}and its role is to take care of the monodromy properties of the form factors as ruled by 
the Watson equations. The equations (\ref{eq:Fmineq}) and the analyticity
requirement of $F_\text{min}$ are able to fix this function up to
normalization, as we will see explicitly for the sh-G model. The
factors $K_n(\th_1,\dots,\th_n)$ in (\ref{generalsolutionW}) must contain all the expected kinematical poles. In
addition, they must fulfill the proper asymptotic behavior. This yields the following final parametrization of the generic
$n$-particle form factor
\be
F_n(\th_1,\dots,\th_n)=H_n\,Q_n(x_1,\dots,x_n)\,\prod^n_{i<j}\frac{F_\text{min}(\th_{ij})}{x_i+x_j}\,,
\label{F_n}
\ee
where $H_n$ is a normalization factor, $x_i=e^{\th_i}$ and
$Q_n(x_1,\dots,x_n)$ is a symmetric polynomial. In view of
Eq.~\erf{eq:asympt}, the polynomial $Q_n$ of a scalar operator has
the total degree equal to the degree of the polynomial $\prod_{i < j}
(x_i+x_j)$ in the denominator, $n(n-1)/2$. The actual expression
of the polynomials $Q_n$ can be determined by solving the recursive
equations (\ref{reskin}). To this aim, it is convenient to make use of
the basis given by the elementary symmetric polynomials
$\sigma^{(n)}_{k}$ of the $n$ variables $x_i$ defined by
\be 
\prod_{i=1}^n(x+x_i)=\sum_{k=1}^n
x^{n-k}\sigma^{(n)}_k(x_1,\dots,x_n)\,, 
\ee or explicitly 
\be
\sigma^{(n)}_k=\sum_{i_1<\dots<i_k}^n x_{i_1}\dots x_{i_n}\,.  
\labl{eq:sigma}

It is worth remarking that there is a one-to-one correspondence
between the infinite set of form factors
$\{F_n(\th_1,\dots,\th_n)\,,\,\,n\in\mathbb{N}\}$ which are the
solutions of the functional and recursive equations, and the operator
content of a relativistic field theory (see \cite{GMJC}). Let us see how this correspondence
is realized in the sh-G model by solving the form factor equations of
this theory.

\subsection{Form factors of the Sinh--Gordon model}
\noindent
For the sh-G model the minimal form factor $F_\text{min}(\th)$, 
solution of  
equations (\ref{eq:Fmineq}), is \cite{FMS}
\begin{equation}
F_\text{min}(\th)={\cal N}\,\exp\left\{4\int_0^\infty\frac{\ud t}t\,
\frac{\sinh\left(\frac{t}2\,\alpha\right)\sinh\left(\frac{t}2(1-\alpha)\right)}
     {\sinh(t)\cosh\left(\frac{t}2\right)}
     \,\sin^2\left(\frac{t\hat\th}{2\pi}\right)\right\}\,,
\label{FMINSHG}
\end{equation}
where $\hat\th=i\pi-\th$ and the normalization constant ${\cal
N}=F_\text{min}(i\pi)$ is chosen to be
\[
{\cal N}=\exp\left\{-2\int_0^\infty\frac{\mathrm{d} t}t\,
\frac{\sinh\left(\frac{t}2\,\alpha\right)\sinh[\frac{t}2(1-\alpha)]}
    {\sinh(t)\cosh\left(\frac{t}2\right)}\right\}=
    \frac1{\cos\left(\frac{\pi\alpha}2\right)}
    \exp\left\{-\frac1\pi\int_0^{\pi\alpha}\ud t\,\frac{t}{\sin(t)}\right\}\,.
\] 
In addition to the functional equations \erf{eq:Fmineq}, $F_\text{min}(\th)$ also satisfies 
\be
F_\text{min}(i\pi+\th)F_\text{min}(\th)=\frac{\sinh\th}{\sinh\th+\sinh(i\pi\alpha)}\,\,\,.
\labl{eq:Fminrel}
With the choice 
\begin{align}
H_{2n+1}&=H_1 \left(\frac{4\sin(\pi\alpha)}{\cal N} \right)^n\,, \nonumber\\
H_{2n}&=H_2 \left(\frac{4\sin(\pi\alpha)}{\cal N} \right)^{n-1}\,,\nonumber 
\end{align}
the recursive equations for the polynomials $Q_n$ 
entering (\ref{F_n}) can be written as 
\be
(-1)^nQ_{n+2}(-x,x,x_1,\dots,x_n)=x D_n(x;x_1,\dots,x_n)Q_n(x_1,\dots,x_n)\,,
\labl{eq:rec}
where the functions $D_n$ are given by  
\be
D_n(x;x_1,\dots,x_n)=\sum_{k=1}^n\sum_{m=1,3,5,\dots}^k(-1)^{(k+1)}[m] x^{2(n-k)+m}\sigma^{(n)}_k\sigma^{(n)}_{k-m}\,.
\ee 
In this formula $\sigma_k^{(n)}$ are the elementary symmetric polynomials \erf{eq:sigma} while  
\be
[k] \,\equiv\,\frac{\sin(k\pi\alpha)}{\sin(\pi\alpha)} \,\,\,.
\ee
As shown in \cite{mussardo}, a solution of the recursive equations 
(\ref{eq:rec}) is given by the class of symmetric polynomials 
\be
Q_n(k)=\det M_n(k)\,,
\ee
where $M_n(k)$ is a $(n-1)\times(n-1)$ matrix with elements
\be
\left[M_n(k)\right]_{i,j}=\sigma^{(n)}_{2i-j}[i-j+k]\,.
\ee
The corresponding form factors are the matrix elements of a continuous
family of operators identified with the exponential fields $e^{kg\phi}$
\cite{mussardo,brazluk}. With the normalization given by 
$H_n(k)=\left(\frac{4\sin(\pi\alpha)}{\cal N} \right)^{n/2}\,[k]$,  
the explicit form of all form factors of these operators is then 
\be
F_n(k)=\langle0| e^{kg\phi}|\th_1,\th_2,\dots,\th_n\rangle =
[k]\left(\frac{4\sin(\pi\alpha)}{{\cal N}}\right)^{\frac{n}2}\det
M_n(k)\prod^n_{i<j}\frac{F_\text{min}(\th_i-\th_j)}{x_i+x_j}\,.
\labl{eq:FFexp}
So, for instance, the one and two-particle form factors are given by 
\begin{align}
\langle0|e^{kg\phi}|\th\rangle &=  \frac2{\sqrt{{\cal N}}}\,
\frac{\sin(k\pi\alpha)}{\sqrt{\sin(\pi\alpha)}}\,,\\ 
\langle0|e^{kg\phi}|\th_1,\th_2\rangle &=  \frac{4}{{\cal N}}  
\frac{\sin^2(k\pi\alpha)}{\sin(\pi\alpha)}\,F_\text{min}(\th_1-\th_2)\,.
\end{align}

From now on we will concentrate our attention on the form factors of
the even powers of the field $\phi$ because we will need only these
operators for the future computation of the expectation values of the
LL model (only even operators can have non-zero expectation
values). It is useful to express the operator content of the theory in
terms of a class of particular operators, denoted by $\no{\phi^k}$,
which start creating $n$ particles out of the vacuum only when $n \geq k$:
\begin{equation}
F^{\no{\,\phi^k\,}}_n(\th_1,\dots,\th_n) \,= \,
0 \,\,\,\,\,\,  \text{if} \,\,\,\,\,\, n < k \,.
\label{previous}
\end{equation} 
In their form factors for $n = k$ the polynomial term
$Q_{2k}(x_1\dots,x_{2k})$ is equal to the polynomial $\prod_{i <
j}^{2k} (x_i + x_j)$ of the denominator and they cancel each other,
giving
\be
F^{\no{\,\phi^k\,}}_k(\th_1,\dots,\th_k) = 
2^{k}k!\left(\frac{\pi^2\alpha^2}{{\cal N} g^2\sin(\pi\alpha)}\right)^{\frac{k}2}\,\prod_{i<j}^k F_\text{min}(\th_{ij})\,.
\label{eq:Fmm}
\ee
In view of the recursive equations (\ref{reskin}), the absence of
kinematical poles in $F^{\no{\,\phi^k\,}}_k(\th_1,\dots,\th_k)$
obviously implies the vanishing values (\ref{previous}).  To
compute the form factors of these operators when $n > k$, we can take
advantage of the knowledge of the form factors (\ref{eq:FFexp}) of the
exponential operators. Let us denote by $\tilde{\phi^m}$ the operator
whose form factors $\tilde F^m_n$ are obtained by extracting the $\mc
O(k^m)$ term in the expansion of $F_n(k)$. In view of equations
(\ref{previous}) and (\ref{eq:Fmm}) we have
\be
\tilde F^{k}_n= F^{\no{\,\phi^{k}\,}}_n + \sum_{l=2,4,\dots}^{k-2}A^{k}_{l}\,F^{\no{\,\phi^{l}\,}}\,,
\label{A_k_l}
\ee
which implies a mixing among the operators $\no{\phi^{k}}$
\begin{align}
\label{eq:mixing}
\tilde{\phi^2}&=\no{\phi^2}\,,\nonumber\\
\tilde{\phi^4}&=\no{\phi^4}+A^4_2\no{\phi^2}\,,\nonumber\\
&\;\,\vdots\nonumber\\
\tilde{\phi^{k}}&=\no{\phi^{k}}+\sum_{l=2,4,\dots}^{k-2}A^{k}_{l}\no{\phi^{l}}\,.
\end{align}
In Appendix A we discuss how to compute iteratively the coefficients 
$A^{k}_{l}$.

\subsection{LeClair--Mussardo formalism}
\label{sec:Tcorr}
\noindent
At equilibrium the expectation value of a local operator $\mc O(x,t)$ at temperature $T$ and at finite density $n$ is given by 
\be
\vev{\mc O}_{T,n} = \frac{\mathrm{Tr}\left(e^{-\frac{H-\mu N}{\kb T}}\mc{O}\right)}
{\mathrm{Tr}\left(e^{-\frac{H-\mu N}{\kb T}}\right)}\,\,.
\labl{eq:Ovev}
For translation invariance at equilibrium $\vev{\mc O}_{T,n}$ is independent of $x$ and $t$. 
If we specify this formula to an integrable quantum field theory and 
we use the basis of
multiparticle scattering states, we have 
\begin{equation}
\vev{\mc O}_{T,n} =
\frac1{Z_{T,n}}\sum_{k=0}^\infty\frac1{k!}
\int_{-\infty}^\infty\frac{\ud\th_1}{2\pi}\dots\frac{\ud\th_k}{2\pi}
\left(\prod_{i=1}^k e^{-\frac{M\cosh\th_i - \mu}{\kb T}}\right) 
\3pt{\th_k,\dots,\th_1}{\mc O(0,0)}{\th_1,\dots,\th_k}\,,
\label{eq:vevorig}
\end{equation}
where $Z_{T,n}=\mathrm{Tr}\left(e^{-\frac{H-\mu N}{\kb T}}\right)$. 
As shown in \cite{leclair}, this expression can be neatly written as 
\be
\vev{\mc O}_{T,n}=\sum_{k=0}^\infty\frac1{k!}
\int_{-\infty}^\infty \frac{\ud\th_1}{2\pi}\dots\frac{\ud\th_k}{2\pi}
\left(\prod_{i=1}^k\frac1{1+e^{\eps(\th_i)}}\right) 
\3pt{\th_k,\dots,\th_1}{\mc O(0,0)}{\th_1,\dots,\th_k}_\text{conn}\,\,\,,
\labl{eq:muss}
where $\eps(\th)$ is the pseudo-energy, solution of the
Thermodynamical Bethe Ansatz equation of the model of interest, while
the connected form factor is defined as \cite{balog}
\begin{equation}
\3pt{\th_k,\dots,\th_1}{\mc O}{\th'_1,\dots,\th'_k}_\text{conn}=
{\cal F}\left(\lim_{\eta_i\to0} \3pt{0}{\mc
  O}{\th'_1,\dots,\th'_k,\th_k-i\pi+i\eta_k,\dots,\th_1-i\pi+i\eta_1}\right)\,,
\label{eq:conndef}
\end{equation}
where ${\cal F}$ in front of the expression means taking its finite part,
that is, omitting all the terms of the form $\eta_i/\eta_j$ and
$1/\eta_i^p$ where $p$ is a positive integer. In Appendix
\ref{sec:Fexpl} we give an explicit example for the calculation of the
connected limit. In this formulation the $\mu$- and $T$-dependence of the right hand
side of \erf{eq:muss} is hidden in $\eps(\th)$ that, for the sh-G
model, satisfies the $\mu$- and $T$-dependent equation
\erf{eq:TBA}. Expression \eqref{eq:muss} was checked in various
cases \cite{saleur,castro} and was compared with the direct
evaluation of the expectation value \erf{eq:Ovev} using finite volume
regularization \cite{pozsgay}.

Notice that for the sh-G model, in view of the functional relation
\erf{eq:Fminrel} the connected limit (\ref{eq:conndef}) for the
product of the $F_{\text{min}}(\th_{ij})$ in the form factors of ${\mc
O}$ simply becomes
\begin{eqnarray}
\prod_{i<j}^{2k}F_\text{min}(\th_{ij}) & \longrightarrow & 
\left(F_\text{min}(i\pi)\right)^k
\prod_{i<j}^k\frac{\sinh\th_{ij}}{\sinh\th_{ij}+\sinh(i\pi\alpha)}\,\frac{\sinh\th_{ji}}{\sinh\th_{ji}+\sinh(i\pi\alpha)}= \noindent \nonumber\\
&=& {\cal N}^k\prod_{i<j}^k \frac{\sinh^2\th_{ij}}{\sinh^2\th_{ij}+\sinh^2(\pi\alpha)}\,.
\label{eq:fmincon}
\end{eqnarray}
This means that it is not necessary to employ the explicit form
(\ref{FMINSHG}) of $F_\text{min}(\th)$ to calculate the connected form
factors and, for their actual determination, we only have to take the
connected limit of the rest of the form factor formula. In particular,
the connected form factor $F^{\no{\,\phi^{2k}\,}}_{2k,\text{conn}}$
can be calculated from the explicit formula \erf{eq:Fmm} for
$F^{\no{\,\phi^{2k}\,}}_{2k}$. Since it only depends on the rapidities
through the $F_\text{min}$ factors, using \erf{eq:fmincon} we can
immediately write down the connected form factor:
\be
F^{\no{\,\phi^{2k}\,}}_{2k,\text{conn}}= 
2^{2k}(2k)!\left(\frac{\pi^2\alpha^2}{g^2\sin(\pi\alpha)}\right)^k\,
\prod_{i<j}^k \frac{\sinh^2\th_{ij}}{\sinh^2\th_{ij}+\sinh^2(\pi\alpha)}\,.
\labl{eq:Fcmax}

\section{The double limit of the Sinh--Gordon model}
\label{sec:limit}
\noindent
In this section we analyze in detail the mapping between the sh-G and
the LL models. We show that it is possible to obtain the LL model from
the sh-G model by taking the non-relativistic limit simultaneously
with the limit $g\to0$, where $g$ is the sh-G coupling constant. In
particular, we show how this mapping is realized at the level of the
$S$-matrix, the Lagrangian densities and the Thermodynamical Bethe
Ansatz equations.

\subsection{Double limit of the two-particle $S$-matrix}
\noindent
Let us consider the exact $S$-matrix of the sh-G model 
\be
S_{\text{sh-G}}(\th,\alpha)=\frac{\sinh\th-i\,\sin(\alpha\pi)}{\sinh\th+i\,\sin(\alpha\pi)}\,\,\,,
\ee
and let us take its non-relativistic limit accompanied by a simultaneous limit of the coupling constant $g$ toward smaller values such that  
\be
c\to\infty\,\,\, ,\,\,\, g\to0,\;\quad g\,c=\text{fixed}\,\,\,.
\label{eq:limit}
\ee
The resulting expression 
\be
S(\th,\alpha) \longrightarrow
\frac{\frac{p}{Mc}-\frac{i\hbar}8g^2c}{\frac{p}{Mc}+\frac{i\hbar}8g^2c}\,
\ee
coincides with the LL $S$-matrix (\ref{SmatrixLLM}) once we set the
sh-G and LL masses equal, $M=m$, and   
\be
\lambda\equiv \frac{\hbar^2c^2}{16}\,g^2\,.
\label{eq:id}
\ee
Hence the $S$-matrices of the two models coincide in this double limit.
It is worth noticing that the resulting coupling $\lambda$ of the LL
model does not need to be small and therefore we shall be able to
study the LL model at arbitrarily large values of its coupling. To use this correspondence between the two models to calculate
correlation functions in the LL model, we need to establish the
relation between the operators of these two theories. For this reason
in the next section we show how to perform the limit \erf{eq:limit} on
the fields and the Hamiltonians.

\subsection{Non-relativistic limit at the Lagrangian level}
\label{sec:LLlimit}
\noindent
Consider the sh-G Lagrangian density
\be
\mc{L}= \frac12\left(\frac{\p\phi}{c\,\p t}\right)^2-\frac12\left(\frac{\p\phi}{\p x}\right)^2-
\frac{m_0^2c^2}{g^2\hbar^2}\left(\cosh(g\,\phi)-1\right)\,.
\labl{eq:Lsg}
To study its non-relativistic limit, it is convenient to write initially the real
scalar field in the form \cite{beg,bergman,dimock,jia} 
\be
\phi(x,t)=\sqrt{\frac{\hbar^2}{2m_0}}\left(\psi(x,t)\,
  e^{-i\frac{m_0c^2}\hbar\,t}+\psid(x,t) e^{+i\frac{m_0c^2}\hbar\,t}\right)\,. 
\labl{eq:phipsi}
Substituting this expression into the Lagrangian (\ref{eq:Lsg}) and taking the 
limit $c\to\infty$, we can discard all the oscillating terms,
that is, terms containing factors $e^{i\,nm_0c^2/\hbar\,t}$ (with $n$ non-vanishing 
positive or negative integers). 
These terms, in fact, oscillate very rapidly in 
this limit and average to zero when integrated over any small but finite
time interval. In more detail, the relativistic canonical momentum can be written as
\begin{multline}
\Pi(x,t)=\frac1{c^2}\dot\phi(x,t)=\\
\shoveleft{\sqrt{\frac{\hbar^2}{2m_0}}\frac1{c^2}\left[\left(\dot\psi(x,t)-\frac{im_0c^2}\hbar\psi(x,t)\right)
  e^{-i\frac{m_0c^2}\hbar\,t}+\left(\dot\psid(x,t)+\frac{im_0c^2}\hbar\psid(x,t)\right)e^{+i\frac{m_0c^2}\hbar\,t}\right]=}\\
=-i\sqrt{\frac{m_0}2}\left(\psi(x,t)\,
  e^{-i\frac{m_0c^2}\hbar\,t}-\psid(x,t) e^{+i\frac{m_0c^2}\hbar\,t}\right)+\mathcal{O}\left(\frac1{c^2}\right)\,.
\end{multline}
This allows us to express $\psi$ and $\psid$ in terms of $\phi$ and
$\Pi$ (up to order $\mathcal{O}\left(\frac1{c^2}\right)$):
\bes
\begin{alignat}{2}
\psi(x,t)&=e^{i\frac{m_0c^2}\hbar}&&\left(\frac1\hbar\sqrt{\frac{m_0}2}\,\phi(x,t)+\frac{i}{\sqrt{2m_0}}\,\Pi(x,t)\right)\,,\\
\psid(x,t)&=e^{-i\frac{m_0c^2}\hbar}&&\left(\frac1\hbar\sqrt{\frac{m_0}2}\,\phi(x,t)-\frac{i}{\sqrt{2m_0}}\,\Pi(x,t)\right)\,.
\end{alignat}
\label{eq:psiphi}
\esu
It is easy to show that the commutation relation
\be
[\phi(x,t),\Pi(x',t)]=i\hbar\,\delta(x-x')
\ee
implies the following commutation relation for the non-relativistic operators
\be
[\psi(x,t),\psid(x',t)]=\delta(x-x')\,.
\ee
Turning to the Lagrangian density, 
the kinetic term $K$ of \erf{eq:Lsg} becomes 
\be
K\longrightarrow \frac{\hbar^2}{2m_0c^2}\frac{\p\psid}{\p t}\frac{\p\psi}{\p t} -
\frac{\hbar^2}{2m_0}\nabla\psid\nabla\psi +
i\,\frac\hbar2\left(\psid\frac{\p\psi}{\p t} - \frac{\p\psid}{\p
  t}\psi\right) + \frac12 m_0c^2\psid\psi\,.
\labl{eq:K}
Expanding the formula \erf{eq:mu_m} for $m_0$ in the combined limit
\erf{eq:id} we obtain
\be
m_0^2 = M^2+\frac23\frac{M^2\lambda^2}{\hbar^2c^2}+\mc O(\frac1{c^4})\,.
\ee
From the second term of \eqref{eq:K} we see again that the physical
masses in the two models should be equal $m=M$ and then the first term
is of order $1/c^2$ and can be dropped in the limit.

Let us now turn our attention to the interaction term $\cosh(g \phi)$
in the Lagrangian (\ref{eq:Lsg}), which is equivalent to an infinite
series in terms of even powers of the field $\phi$. Expressing $\phi$
in terms of the new fields $\psi$ and $\psid$, we can use the binomial
formula to expand each power $\phi^{2k}$ in terms of these
fields. Taking into account that the oscillating terms should be
dropped, only the symmetric ``middle term'' of the binomial expansion
survives from each power. Collecting the combinatorial factors from
the different expansions of the powers, we arrive at the following
series:
\begin{multline}
U(\phi)=\frac{m_0^2c^2}{g^2\hbar^2}\left(\cosh(g\phi)-1\right) =
\frac{m_0^2c^2}{g^2\hbar^2}\sum_{n=1}^\infty \frac1{(2n)!}(g\phi)^{2n}\longrightarrow\\
\longrightarrow \frac{m_0^2c^2}{g^2\hbar^2}\sum_{n=1}^\infty
\frac1{(2n)!}\left(\frac{\hbar^2}{2m_0}\right)^n
\binom{2n}{n}g^{2n}\psid\,^n\psi^n = \sum_{n=1}^\infty \frac1{(n!)^2}
\frac{m_0^2c^2}{g^2\hbar^2} \left(\frac{\hbar^2g^2}{2m_0}\right)^n\psid\,^n\psi^n\,.
\end{multline}
The $n=1$ term of the series,
\be
\frac{m_0c^2}{2}\psid\,\psi\,,
\ee
exactly cancels the last term of \eqref{eq:K}.
The $n=2$ term becomes 
\be
\frac{\hbar^2 c^2g^2}{16}\psid\,^2\psi^2\longrightarrow\lambda\,\psid\,^2\psi^2\,,
\ee
which is just the interaction term in the LL Lagrangian. The rest of the series can be organized as
\be
\sum_{n=3}^\infty \left[\frac1{2^n(n!)^2}\frac{c^2}{m_0^{n-2}} (\hbar^2g^2)^{n-1}\psid\,^n\psi^n + \dots\right]
= \sum_{n=3}^\infty\frac{2^{3n-4}}{(n!)^2}\frac{\lambda^{n-1}}{(mc^2)^{n-2}}\psid\,^n\psi^n+\dots\,,
\ee
where the dots indicate possible higher order terms in $1/c$. If we
now take the limit \erf{eq:limit} all the terms in this series
vanish because $\lambda$ is fixed while $c\to\infty$.

In summary, in the double scaling limit the Lagrangian density \erf{eq:Lsg} of the sh-G model becomes
the Lagrangian density \eqref{LagrangianNLS} of the LL model
\be
\mc{L}\;\longrightarrow\;\mc{L}' = -\frac{\hbar^2}{2m}\nabla\psid\nabla\psi + 
i\,\frac\hbar2\left(\psid\frac{\p\psi}{\p t} - \frac{\p\psid}{\p
  t}\psi\right) - \lambda\,\psid\psid\psi\psi\,.
\ee
So by keeping the coefficient of the $\psi^4$ term fixed, which is the
actual constraint enforced by the double limit \erf{eq:limit}, all the higher order 
terms go to zero and we are left with the non-relativistic LL Hamiltonian.


\subsection{The non-relativistic limit of the Sinh--Gordon TBA equations}
\label{sec:TBAlim}
\noindent 
To study the non-relativistic limit of the TBA equations of
Section \ref{sec:sg} it is convenient to make the coordinate change
(using from now that $m=M$)
\be
p=mc\sinh\th\,\,, \qquad \ud p=mc\cosh\th \,\ud\theta\,.
\ee
Using 
\be
\int_{-\infty}^\infty\ud\th\, \rho^\text{(r)}(\th)=\frac{N}L =
\int_{-\infty}^\infty\ud p\, \tilde\rho^\text{(r)}(p)\,,
\labl{eq:intrho}
this implies
\be
\tilde\rho^\text{(r)}(p)=
\frac1{mc\cosh\th(p)}\,\rho^\text{(r)}\left(\th(p)\right)\approx \frac1{mc}\,\rho^\text{(r)}\left(\frac{p}{mc}\right)
\ee
and 
\be
\tilde\rho(p) \approx \frac1{mc}\,\rho\left(\frac{p}{mc}\right)\,.
\ee
For the sh-G model 
\be
\chi(\th)=-2\arctan\left(\frac{\sin(\alpha\pi)}{\sinh(\theta)}\right)
\Longrightarrow 
 \fii(\th)=\frac{2\sin(\alpha\pi)\cosh(\th)}{\sinh^2(\th)+\sin^2(\alpha\pi)}\,\,\,,
\ee
and in the double limit \erf{eq:limit} the kernel $\fii(\th)$ becomes
\be
\fii(\theta)\longrightarrow mc\,\tilde\fii(p)=
mc\,\frac{4\hbar m\lambda}{\hbar^2p^2+4m^2\lambda^2}\,\,\,. 
\ee
Therefore the TBA equations transform into equations \erf{eq:YYTBA1}
\bes
\begin{align}
2\pi\tilde\rho(p)&=\frac1\hbar + \int_{-\infty}^\infty\ud
p'\,\tilde\fii(p-p')
\,\tilde\rho^{\text{(r)}}(p')\,,\label{eq:YYrhos}\\
\tilde\eps(p)& = -\frac{\tilde\mu}{\kb T}+\frac{p^2}{2m\kb T}-
\int_{-\infty}^\infty\frac{\ud p'}{2\pi}\, 
\tilde\fii(p-p')\log\left(1+e^{-\tilde\eps(p')}\right)\,,\label{eq:YYeps}
\end{align}
\label{eq:YYTBA}
\esu
and $\tilde\rho/\tilde\rho^\text{(r)}=1+e^{\tilde\eps}$, where
\begin{align}
\tilde\eps(p)&=\eps\left(\frac{p}{mc}\right)\,,\\
\tilde\mu&=\mu-mc^2\,.
\end{align}
Observe that it is correct to take the small $p$ and $\theta$ limit in
the integrands even though the integrals are extended to arbitrarily
large momenta: as a matter of fact, the integrals have a finite
support because their integrands vanish asymptotically very fast. The
expressions for the energies become
\bes
\label{eq:YYEs}
\begin{align}
\frac{\tilde E}L=\frac{E-N\,mc^2}L& = 
\int_{-\infty}^\infty\ud p\,\frac{p^2}{2m}\,\tilde\rho^\text{(r)}(p)\,,\\
\frac{\tilde F}L=\frac{F-N\,mc^2}L&=
\tilde\mu n-\frac{\kb T}{2\pi\hbar}\int_{-\infty}^\infty\ud p\, 
\log\left(1+e^{-\tilde\eps(p)}\right)\,,\label{eq:FE}
\end{align}
\esu
which coincide with equations (\ref{eq:YYEs1}).
Similarly, the limit of the $T=0$ equations \erf{eq:TBAT0} is given by 
\bes
\begin{align}
2\pi\tilde\rho^{\text{(r)}}(p)&=\frac1\hbar + \int_{-B}^B\ud
p'\,\tilde\fii(p-p') 
\,\tilde\rho^{\text{(r)}}(p')\,,\label{eq:YYT0rhos}\\
\tilde\eps_0(p)& = -\tilde\mu+\frac{p^2}{2m}+
\int_{-B}^B\frac{\ud p'}{2\pi}\, 
\tilde\fii(p-p')\,\tilde\eps_0(p')\,,
\end{align}
\label{eq:YYT0}
\esu
once again in agreement with equations \erf{eq:YYT01}.

We saw at the end of Section \ref{sec:sg} that the
pseudo-energy describes the dressed energy of the excitations of the
system, which is given by
\be
E(\th)=\kb T \eps(\th)+\mu
\ee
for the sh-G model and by
\be
\tilde E(\th)=\kb T \tilde\eps(p)+\tilde\mu
\ee
for the LL model. It is worth observing the different behaviors of the
excitation energies in the two models. As it can be seen from the TBA
equation \erf{eq:TBA}, the sh-G energy has a gap $M$ that implies that
the correlation functions decay exponentially. On the contrary, the LL
excitation energy starts as $p^2/2m$ for small momenta (see
Eq.~\erf{eq:YYeps}), implying a power-law decay for the correlation
functions.  Our double limit takes care of this difference
automatically and thus it will give correct results for the LL model.

\section{Local correlators for the \LL model}
\label{sec:LL1pt}
\noindent 
In this section we calculate LL one-point correlation functions at
fixed particle density $n$ and temperature $T$ by applying the formulas 
of Sections \ref{sec:models} and \ref{sec:FF}, in connection with the double limit presented in Section \ref{sec:LLlimit}. The fields are
taken at the same position and time: since our system is 
taken at equilibrium and translationally invariant, their correlators are obviously space and time independent. 
We focus our attention on the local $k$-particle correlation 
functions $g_k$ defined as 
\be
\vev{\psid\,^k\psi^k}=n^k\,g_k(\gamma,\tau)\,\,\,,
\ee
where $\gamma$ and $\tau$ are given in \erf{eq:gamma} and \erf{eq:tau}.
These local correlators play an important role in experiments 
with ultracold bosons since the pair correlations are responsible 
for the rates of inelastic collisional processes. Furthermore, 
the low-temperature recombination rate for a Bose gas is proportional 
to the local three-body correlation function \cite{kagan} and measurements of 
the three-body recombination rate can be used to determine the local correlations 
and as a tool for distinguishing condensed and non-condensed phases 
\cite{cornell}. 
For $g_2(\gamma,\tau)$ and $g_3(\gamma,\tau=0)$ exact results are
available \cite{gangardt,kheruntsyan,cheianov}, whereas for the others
the asymptotic behavior of the correlators in the regimes of small
and large coupling or temperature was computed in
\cite{gangardt,kheruntsyan}. We will present a comparison 
with the exact and the approximate results present in the literature, showing 
the improvement that our method brings in the computation of these 
correlators.  

Let us start with the correspondence between the sh-G and the LL operators 
\be
\vev{\no{\phi^{2k}}}\longrightarrow \left(\frac{\hbar^2}{2m}\right)^k\binom{2k}{k}\vev{\psid\,^k\psi^k}\,,
\labl{eq:natcorr}
which can be established along the limit procedure described in
Section \ref{sec:LLlimit}.  To compute the expectation values of these
operators at finite density $n$ of the Bose gas and at finite
temperature, we need to employ the results of Section \ref{sec:FF}, so
that
\begin{equation}
\vev{\psid\,^k\psi^k} = 
\lim\, \binom{2k}{k}^{-1}\!\!\left(\frac{\hbar^2}{2m}\right)^{-k}\sum_{l=k}^{\infty}\frac1{l!}\int_{-\infty}^{\infty}\frac{\ud
\th_1}{2\pi}\,f(\th_1)\dots \int_{-\infty}^{\infty}\frac{\ud
\th_l}{2\pi}\,f(\th_l)\,F^{\no{\,\phi^{2k}\,}}_{2l}(\th_1,\dots,\th_l)_\text{conn}\,,
\label{expr_arr}
\end{equation}
where $f(\th)=1/(1+e^{\eps(\th)})$ are the filling fractions and the notation ``$\lim$'' 
denotes the double limit \erf{eq:limit}. Note that 
in Eq.~(\ref{expr_arr}) 
the terms with $l<k$ are zero and therefore the first non-zero term in
the series is a $k$-fold integral. Now, similarly to what happens for the
TBA equations, the filling fractions effectively cut off the integrands at large 
values of the rapidities, so we can exchange the order of the limit and the 
integrals, arriving at a fully non-relativistic formula
\begin{equation}
\vev{\psid\,^k\psi^k} = 
\binom{2k}{k}^{-1}\!\!\left(\frac{\hbar^2}{2m}\right)^{-k}\sum_{l=k}^{\infty}\frac1{l!}\int_{-\infty}^{\infty}\frac{\ud
p_1}{2\pi}\,f(p_1)\dots \int_{-\infty}^{\infty}\frac{\ud
p_l}{2\pi}\,f(p_l)\,\tilde F^{\no{\,\phi^{2k}\,}}_{2l}(p_1,\dots,p_l)_\text{conn}\,.
\label{eq:formula}
\end{equation}
Here $f(p)=1/(1+e^{\tilde\eps(p)})$ where $\tilde\eps(p)$ is the solution of the non-relativistic TBA equations
(\ref{eq:YYTBA},\ref{eq:intrho}) and
\be
\tilde
F^{\no{\,\phi^{2k}\,}}_{2l}(\{p_i\})_\text{conn}=\lim\,\left(\frac1{mc}\right)^lF^{\no{\,\phi^{2k}\,}}_{2l}(\{\th_i=\frac{p_i}{mc}\})_\text{conn}
\ee
are the double limit of the connected form factors. We go through the
steps of the calculation of a specific form factor and we list the
explicit expressions of the first few of them in Appendix \ref{sec:Fexpl}. 

A first check of the validity of formula \erf{eq:formula} is provided by the correlator 
$\vev{\psid(x,t)\psi(x,t)}$. With the explicit connected form factors
of $\no{\phi^2}$ the first terms read
\begin{multline}
\vev{\psid\psi}=
\int_{-\infty}^\infty\frac{\ud p}{2\pi}f(p)\frac1\hbar \,+
\int_{-\infty}^\infty\frac{\ud p_1}{2\pi}\int_{-\infty}^\infty\frac{\ud
  p_2}{2\pi}f(p_1)f(p_2)\frac1\hbar\,\tilde\fii(p_{12}) \,\\
  +\int_{-\infty}^\infty\frac{\ud p_1}{2\pi}\int_{-\infty}^\infty\frac{\ud
  p_2}{2\pi}\int_{-\infty}^\infty\frac{\ud p_3}{2\pi}
  f(p_1)f(p_2)f(p_3)\frac1\hbar\,\tilde\fii(p_{12})\tilde\fii(p_{23}) +\dots  \,,
\label{eq:nrecurs}
\end{multline}
where we use the notation $p_{ij}=p_i-p_j$ and $\tilde\fii(p)$ is the
scattering phase shift in the \LL model
\erf{eq:fiiLL}. The pattern in \erf{eq:nrecurs} persists for the
further multiple integrals and one easily recognizes that the right
hand side of \erf{eq:nrecurs} is nothing else but the recursive
expansion of
\be
n=\int_{-\infty}^\infty\ud p\, \tilde\rho^\text{(r)}(p)\,,
\ee
where $\tilde\rho^\text{(r)}(p)$ is the
iterative solution of the integral equation \erf{eq:YYrhos}
\be
f^{-1}(p) \tilde\rho^{\text{(r)}}(p)=\frac1{2\pi\hbar} + \int_{-\infty}^\infty\frac{\ud p'}{2\pi}\,\tilde\fii(p-p') \,\tilde\rho^{\text{(r)}}(p')\,.  
\ee
In this way we successfully recover the identity $\vev{\psid\psi}=n$.
While the result may appear obvious, it is worth stressing that it was
obtained by taking the double scaling limit of the sh-G form
factor expansion and employing the LeClair--Mussardo formalism, so it
provides an important check of the method.

\subsection{Correlators at $T=0$}
\noindent 
At zero temperature, similarly to the TBA equations, the formula \erf{eq:formula} 
for the expectation values admits a simpler expression
\be
\vev{\psid\,^k\psi^k}= \\
\binom{2k}{k}^{-1}\!\!\left(\frac{\hbar^2}{2m}\right)^{-k}\sum_{l=k}^{\infty}\frac1{l!}\int_{-B}^{B}\frac{\ud
p_1}{2\pi}\dots \int_{-B}^{B}\frac{\ud
p_l}{2\pi}\,\tilde F^{\no{\,\phi^{2k}\,}}_{2l}(p_1,\dots,p_l)_\text{conn}\,,
\labl{eq:formulaT0}
where $B$ is the Fermi momentum determined by the set of TBA equations
\erf{eq:YYT0} together with
\be
n=\int_{-B}^B\tilde\rho^\text{(r)}(p)\,\ud p \,.
\labl{eq:nint}
The equations become more transparent by introducing the dimensionless quantities
\be
k \equiv \frac{p}B\,,\quad \nu(k)\equiv\hbar\tilde\rho^\text{(r)}(B k)\,, \quad \beta \equiv \frac{2m}\hbar\frac\lambda{B}=\frac{\hbar n\gamma}B\,,
\ee
where in the last expression we used the definition of the LL
parameter $\gamma$ of Eq.~\erf{eq:gamma}. In terms of these new variables, 
equations (\ref{eq:nint},\ref{eq:YYT0rhos}) become 
\bes
\begin{align}
1&=\frac\gamma\beta\int_{-1}^1\nu(k)\,\ud k\,,\label{eq:betgam}\\
\nu(k)&=\frac1{2\pi} + \int_{-1}^1\frac{\ud k'}{2\pi}\,\frac{2\beta}{(k-k')^2+\beta^2}\,\nu(k')\,,\label{eq:f_inteq}
\end{align}
\label{eq:dimlT0}
\esu
while the (non-relativistic) ground state energy is given by
\be
\frac{\tilde E}L=\int_{-B}^B\ud p\,\tilde\rho^\text{(r)}(p)\frac{p^2}{2m} =
\frac{\hbar^2}{2m}\,n^3\left(\frac\gamma\beta\right)^3\int_{-1}^1\ud
k\,\nu(k)k^2\equiv \frac{\hbar^2}{2m}\,n^3\,e(\gamma)\,. 
\labl{eq:e}
The strong coupling expansion of \erf{eq:betgam} is obtained by plugging into it the iterative solution of \eqref{eq:f_inteq} and expanding the integrals in $\beta^{-1}$ 
\begin{multline}
1=\left(\frac{\gamma}{\beta}\right)\left\{\frac1\pi+\frac2{\pi^2}\beta^{-1}(1-\frac23\beta^{-2}+\frac{16}{15}\beta^{-4}+\dots)\right.\\
+\left.\frac4{\pi^3}\beta^{-2}(1-\frac43\beta^{-2}+\frac{8}{3}\beta^{-4}+\dots)
+ \frac8{\pi^4}\beta^{-3}(1-2\beta^{-2}+\dots) +\dots\right\}\,.
\label{eq:g2exp}
\end{multline}
This provides a series expansion relation between $\beta$ and $\gamma$ 
\be
\gamma=\pi\beta-2+\frac4{3\beta^2}+\dots\quad \Longleftrightarrow\quad
\beta=\frac1\pi\left(\gamma+2-\frac{4\pi^2}{3\gamma^2}  +\dots \right)\,,
\labl{eq:betagamm}
which is equivalent to a Fermi momentum
\be
B=\hbar n\pi \left(1-\frac2\gamma+\frac4{\gamma^2}+\dots\right)\,.
\labl{eq:B-n}
Using the formulas above,  we can now derive the leading order behavior in
$\gamma^{-1}$ of the general correlator $g_k(\gamma)$. It is easy 
to see that the leading order comes from the first non-zero
integral in the series \erf{eq:formulaT0} where the integrand is the
double limit of the connected form factor \erf{eq:Fcmax}. Taking the double limit, 
for the first term of $\vev{\psid\,^k\psi^k}$ we get
\begin{multline}
\vev{\psid\,^k\psi^k} = 
\left(\frac{\hbar^2}{2m}\right)^{\!\!-k}\!\!\binom{2k}{k}^{-1}\frac1{k!}\int_{-B}^B\frac{\ud p_1}{2\pi}\dots\int_{-B}^B\frac{\ud p_k}{2\pi}
2^{2k}(2k)!\left(\frac\hbar{8m}\right)^k\prod_{i<j}^k\frac{\hbar^2 p_{ij}^2}{\hbar^2 p_{ij}^2+4m^2\lambda^2}+\cdots \\
=\, \left(\frac{B}\hbar\right)^k \!\!\frac{k!}{(2\pi)^k}\int_{-1}^1\ud k_1\dots\int_{-1}^1\ud k_k\prod_{i<j}^k\frac{k_{ij}^2}{k_{ij}^2+\beta^2}+\cdots\,.
\end{multline}
For the leading order behavior of these quantities, we need to keep
only the leading order term $\mc O(\beta^{-n(n-1)})$ of the integrand
and substitute from \erf{eq:B-n} and
\erf{eq:betagamm} $B=\hbar n\pi$ and $\beta=\gamma/\pi$. The final result is 
\be 
g_k=\frac{k!}{2^k}\left(\frac\pi\gamma\right)^{k(k-1)} I_n +\dots\,,
\labl{eq:LO}
where
\be
I_n=\int_{-1}^1\ud k_1\dots\int_{-1}^1\ud k_k \prod_{i<j}^k k_{ij}^2\,.
\ee
So, in this way we recover the expression obtained in \cite{gangardt} by using
completely different methods. Let us now discuss in more detail the
results for $g_1$, $g_2$ and $g_3$. 

\subsubsection{Expectation value $g_1$}
\noindent 
We already demonstrated that our series expansion sums up to the exact
value $g_1=1$, here we show how convergent the series is.  The actual
computation consists of the following steps. First we solve
numerically the integral equation \erf{eq:f_inteq}, second we obtain
the $\beta(\gamma)$ function from \erf{eq:betgam}, and finally, we
integrate numerically the dimensionless forms of the integrals on the
right hand side of \erf{eq:formulaT0}.

\begin{figure}[t]
\centerline{\scalebox{0.3}{\includegraphics{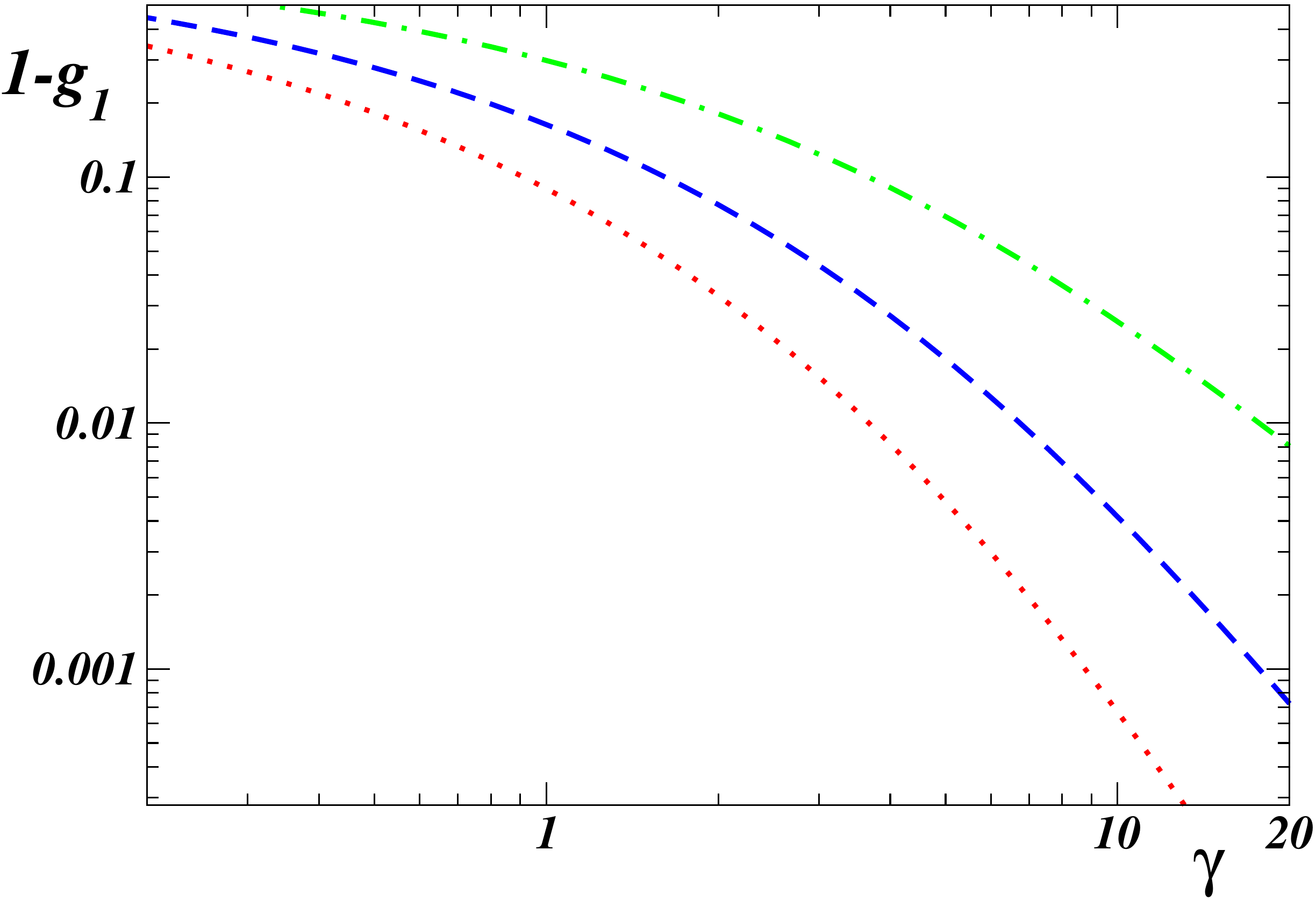}}}
\caption{Deviations $1-g_1$ from the exact result ($g_1=1$) 
at $T=0$ using form factors up to 
$n=\textcolor{green}{4}$, $\textcolor{blue}{6}$ and $\textcolor{red}{8}$ particles, 
respectively with green dot-dashed,  
blue dashed and red dotted lines.}
\label{fig:g1}
\end{figure}

Carrying out these steps for $g_1=\vev{\psid\psi}/n$, we obtain the
plot shown in Fig.~\ref{fig:g1} for $g_1$ as a function of $\gamma$.
The series (\ref{eq:formulaT0}) is nicely saturated by the first few
terms for sufficiently large values of $\gamma$ (one should keep in
mind that $\gamma = 0$ is a singular point of the LL model and
therefore one cannot expect {\it a priori} any fast convergence nearby). In
fact, this plot shows that the exact value $g_1=1$ is
rapidly approached by just the first terms of (\ref{eq:formulaT0}). It
is clear that including more terms in the series, (i.e., employing
higher particle form factors) extends the fast convergence toward
smaller values of $\gamma$. Notice, however, that the convergence of
the series is always remarkably fast for all $\gamma
\geq 1.5$, where the exact value is obtained within a $5\%$ accuracy
just using its first four terms.

\subsubsection{Expectation value $g_2(\gamma)$}
\noindent
Let us continue our discussion with the calculation of the correlation function 
$\vev{\psid\psid\psi\psi}=n^2\,g_2$.
This correlator is the expectation value of the interaction term in
the \LL Hamiltonian \erf{eq:HLL}, thus it can be exactly determined 
\cite{gangardt,kheruntsyan} via the Hellmann--Feynman theorem \cite{HF}:
\be
\vev{\psid\psid\psi\psi}=\frac1L\left<\frac{\ud H}{\ud\lambda}\right>=\frac{\ud
  }{\ud\lambda}\left(\frac{\tilde E}L\right)\;\; \Longrightarrow \;\; g_2=\frac{\ud e(\gamma)}{\ud\gamma}\,,
\labl{eq:HF}
where $e(\gamma)$ was defined in \erf{eq:e}. Our result can be
compared with this expression, providing a good possibility to check again the 
correctness of the approach. We have 
\be
\vev{\psid\psid\psi\psi}=
\int_{-B}^B\frac{\ud p_1}{2\pi}\int_{-B}^B\frac{\ud
  p_2}{2\pi} \frac1{2m\lambda\hbar}\,\tilde\fii(p_{12})p_{12}^2 + 
\int_{-B}^B\frac{\ud p_1}{2\pi}\int_{-B}^B\frac{\ud
  p_2}{2\pi}\int_{-B}^B\frac{\ud p_3}{2\pi}\frac1{2m\lambda\hbar} 
\tilde\fii(p_{12})\tilde\fii(p_{23})p_{13}^2
+\dots
\ee
In terms of the dimensionless quantities the expansion in $\beta$ up
to the four-integral term gives
\begin{multline}
\vev{\psid\psid\psi\psi}=n^2\frac{\gamma^2}{\beta^2}\left\{\frac4{3\pi^2}\beta^{-2}\left(1-\frac85\beta^{-2}+\frac{24}{7}\beta^{-4}+\cdots\right)\right.\\
+\left.\frac8{3\pi^3}\beta^{-3}\left(1-\frac{8}5\beta^{-2}+\frac{332}{105}\beta^{-4}+\cdots\right)
+ \frac{16}{3\pi^4}\beta^{-4}\left(1-\frac{34}{15}\beta^{-2}+\dots\right) +\cdots\right\}\,. 
\end{multline}
Substituting the relation \erf{eq:betagamm} we obtain the
strong coupling expansion
\be
g_2=\frac43\frac{\pi^2}{\gamma^2}\left(1-\frac6\gamma+(24-\frac85\pi^2)\frac1{\gamma^2}\right)+\mc O(\gamma^{-5})\,.
\labl{eq:str-coupl}
The leading order behavior agrees with \erf{eq:LO} but it is worth
noticing that we also obtained subleading terms in $\gamma^{-1}$. We notice that this result can be obtained using the
Hellmann--Feynman theorem and the expansion of the ground-state
energy given in \cite{astra-boro}.

\begin{figure}[t]
\centerline{\scalebox{0.35}{\includegraphics{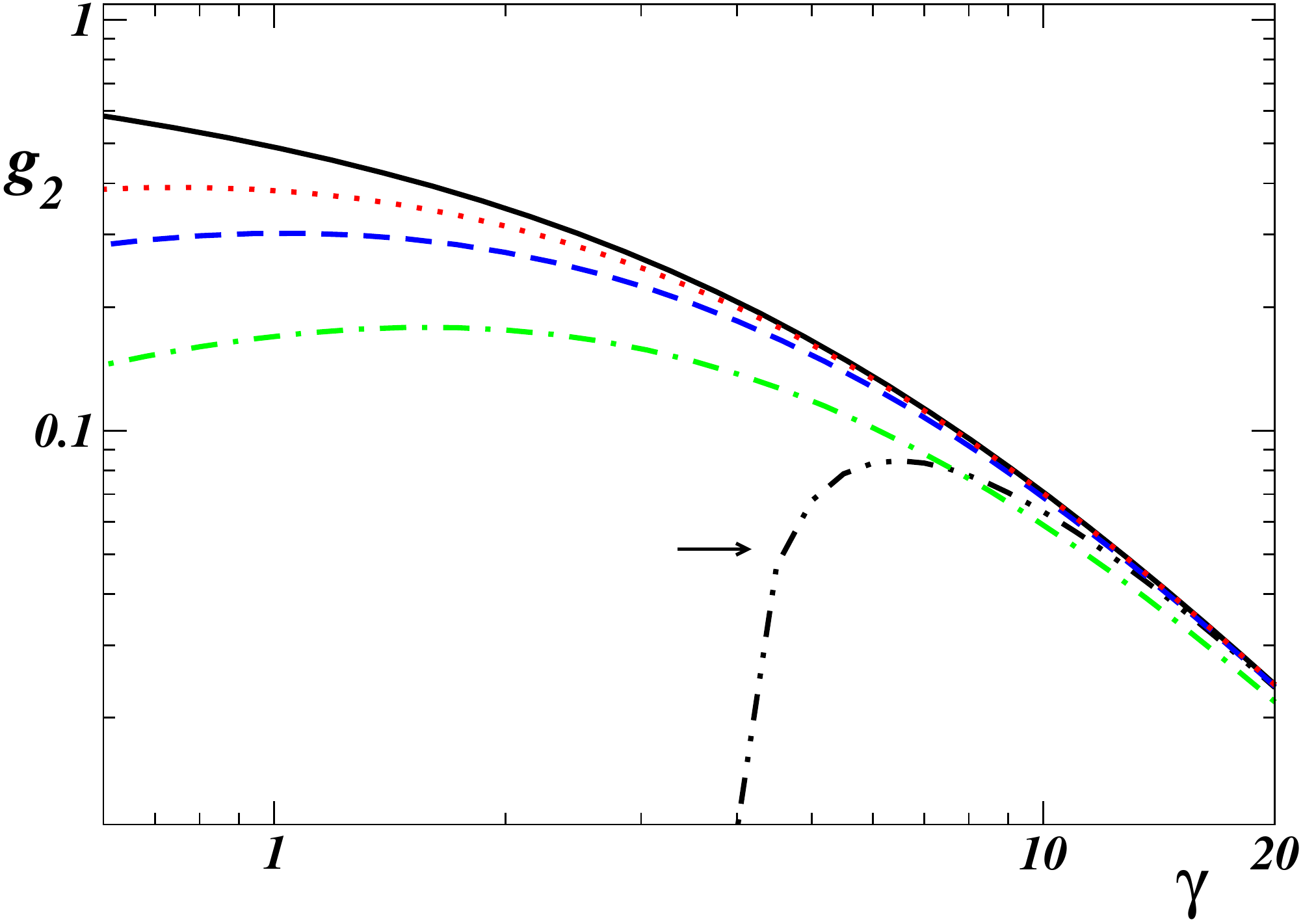}}}
\caption{Plot of $g_2$ as a function of $\gamma$ 
at $T=0$ using form factors up to 
$n=\textcolor{green}{4}$, $\textcolor{blue}{6}$ and $\textcolor{red}{8}$ particles, 
respectively with green dot-dashed,  
blue dashed and red dotted lines. The exact value is given by the 
solid line whereas 
the dot-dot-dashed line below, indicated by the arrow, 
corresponds to the strong coupling 
expansion \erf{eq:str-coupl}.}
\label{fig:g2}
\end{figure}

The plot of $g_2(\gamma)$ -- obtained by numerical integration and
using the integral equations \erf{eq:dimlT0} for $\beta(\gamma)$ -- is
drawn in Fig.~\ref{fig:g2}. As for the previous example, we see that
by increasing the number of form factors employed in our series, our
result rapidly converges to the exact value. The discrepancy between
the exact value and the one obtained with four integrals is less than
$3\%$ for $\gamma>2$. Expression \erf{eq:str-coupl} is also plotted in
Fig.~\ref{fig:g2} to show that the determination of $g_2$ (at
finite $\gamma$) obtained from the first terms of
Eq.~\erf{eq:formulaT0} is much closer to the exact result. This is
because we solve the TBA equations \erf{eq:dimlT0} with an arbitrary
precision and every term of our series contains infinitely many powers
of $\gamma$.

\subsubsection{Expectation value $g_3(\gamma)$}
\noindent 
As a final example let us discuss $g_3$, a quantity known exactly up to now only at
$T=0$ \cite{cheianov}. An analysis similar to the previous cases reveals that
\be
\vev{\psid\psid\psid\psi\psi\psi}=n^3\frac{\gamma^3}{\beta^3}\left\{\frac{16}{15\pi^3}\beta^{-6}\left(1-\frac{144}{35}\beta^{-2}+\cdots\right) + 
\frac{32}{15\pi^4}\beta^{-7}\left(1+\dots\right)+\cdots\right\}\,. 
\ee
Trading $\beta$ for $\gamma$ using \erf{eq:betagamm} we arrive at
\be
g_3=\frac{16}{15}\frac{\pi^6}{\gamma^6}\left(1-\frac{16}\gamma \right)+\mc O(\gamma^{-8})\,.
\labl{eq:g3nlo}
Here the leading order term is the asymptotic result \erf{eq:LO}, but as in the 
previous example, 
we also obtained the next order in the large $\gamma$ expansion.

\begin{figure}[t]
\centerline{\scalebox{0.3}{\includegraphics{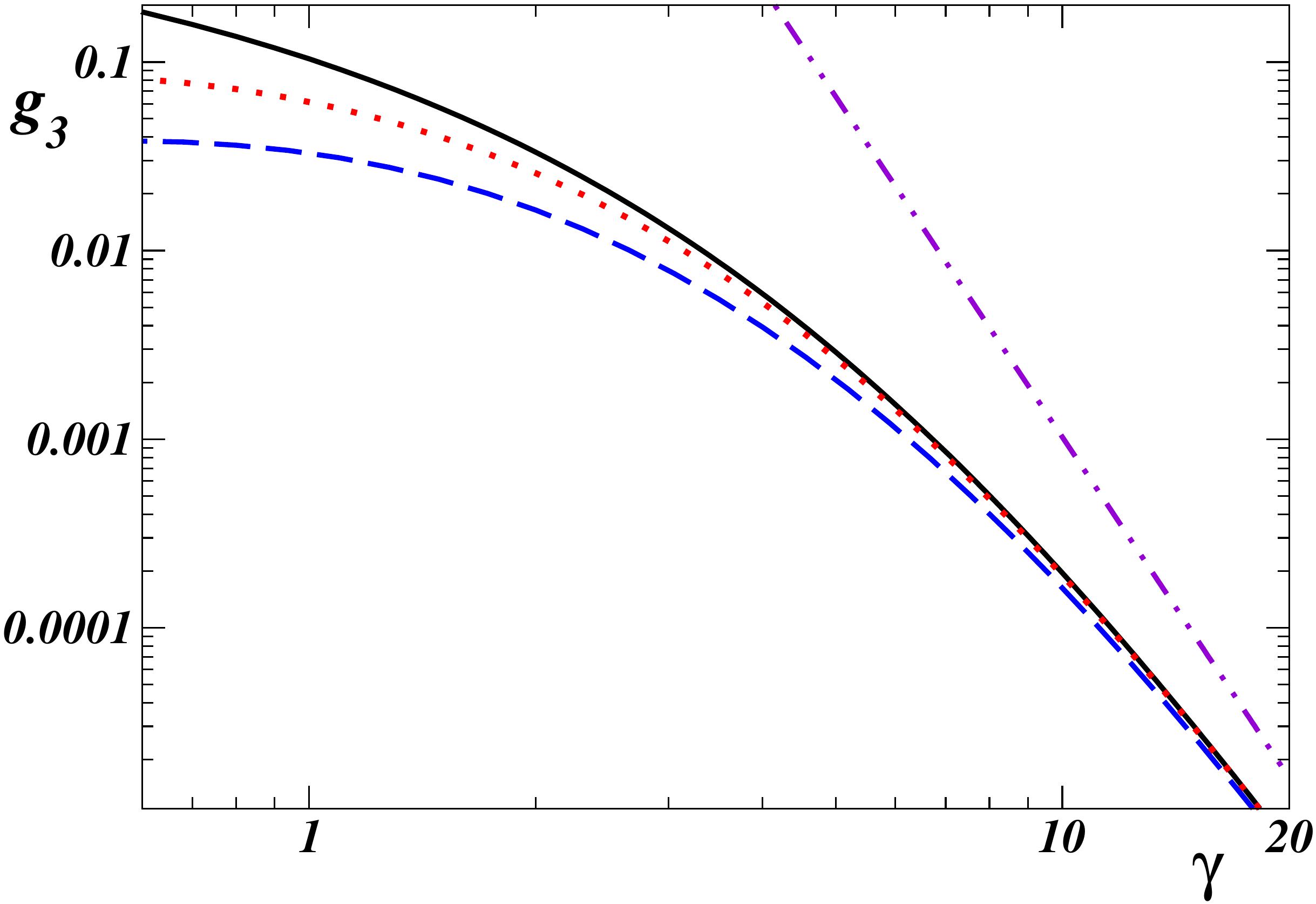}}}
\caption{$g_3$ at $T=0$ with form factors up to $n=\textcolor{blue}{6}$ and $\textcolor{red}{8}$ particles with blue dashed and red dotted lines, 
respectively. The exact value is given by the 
solid line whereas the purple dot-dot-dashed line above corresponds to the
leading order expression \erf{eq:LO}.}
\label{fig:g3}
\end{figure}

The logarithmic plot of $g_3$ using the form factor expansion up to
$n=6$ and 8 particles (one or two terms from the series) is shown in
Fig.~\ref{fig:g3} together with the exact result
of \cite{cheianov}. As in the previous examples, this plot shows a
nice convergent pattern toward the exact value. The leading
order \erf{eq:LO} in the large $\gamma$ expansion is also plotted in
Fig.~\ref{fig:g3} to show that in this domain of $\gamma$
this result largely differs from the exact value. The subleading term,
given in \erf{eq:g3nlo} provides an improvement for larger $\gamma$,
however, the result is still quite far from the one obtained with our
method.

\subsection{Correlators at finite temperature}
\noindent 
To obtain the expectation value at finite values of the
temperature $T$ we have to employ the formula \erf{eq:formula} which contains 
non-trivial filling fractions and we need to solve the whole set of
TBA equations (\ref{eq:intrho},\ref{eq:YYTBA}). It proves to be useful
to introduce now a different set of dimensionless quantities
\be
q \equiv \frac{p}{n\hbar\gamma}\,,\quad
\alpha \equiv \frac{\tilde\mu}{k_\text{B}T}\,,\quad g(q) \equiv 
\frac\hbar\alpha\,\tilde\rho(n\hbar\gamma\,q)\,,
\labl{eq:resc}
which satisfy
\bes
\begin{align}
\tilde\eps(q)& = -\alpha+\frac{q^2\gamma^2}{\tau}-
\int_{-\infty}^\infty\frac{\ud q'}{2\pi}\, 
\frac{2}{(q-q')^2+1}\log\left(1+e^{-\tilde\eps(q')}\right)\,,\label{eq:epsq}\\
g(q)&=\frac1{2\pi\alpha} + \int_{-\infty}^\infty\ud
q'\,\frac{2}{(q-q')^2+1}\,\frac{g(q')}{1+e^{\tilde\eps(q')}}\,,\label{eq:gq}
\end{align}
and
\be
\frac1{\alpha\gamma}=\int_{-\infty}^\infty \,\frac{g(q)}{1+e^{\tilde\eps(q)}}\,\ud q \,.
\ee
\label{eq:diml}
\esu
The physical parameters of the problem are $\lambda$, $T$ and $n$, but
only the dimensionless combinations $\gamma$ and $\tau$ enter the
results. The chemical potential (or the dimensionless fugacity-like
parameter $\alpha$) gets fixed by the constraint given by the last
equation. Rescaling the density with $\alpha$ in \erf{eq:resc}
makes it possible to find the self-consistent solution by an iteration of 
the system of equations.
Once we have $\tilde\eps(q)$ for a given $\gamma$ and $\tau$ we can
substitute it into the form factor expansion \erf{eq:formula}.

\begin{figure}[t]
\centerline{\scalebox{0.3}{\includegraphics{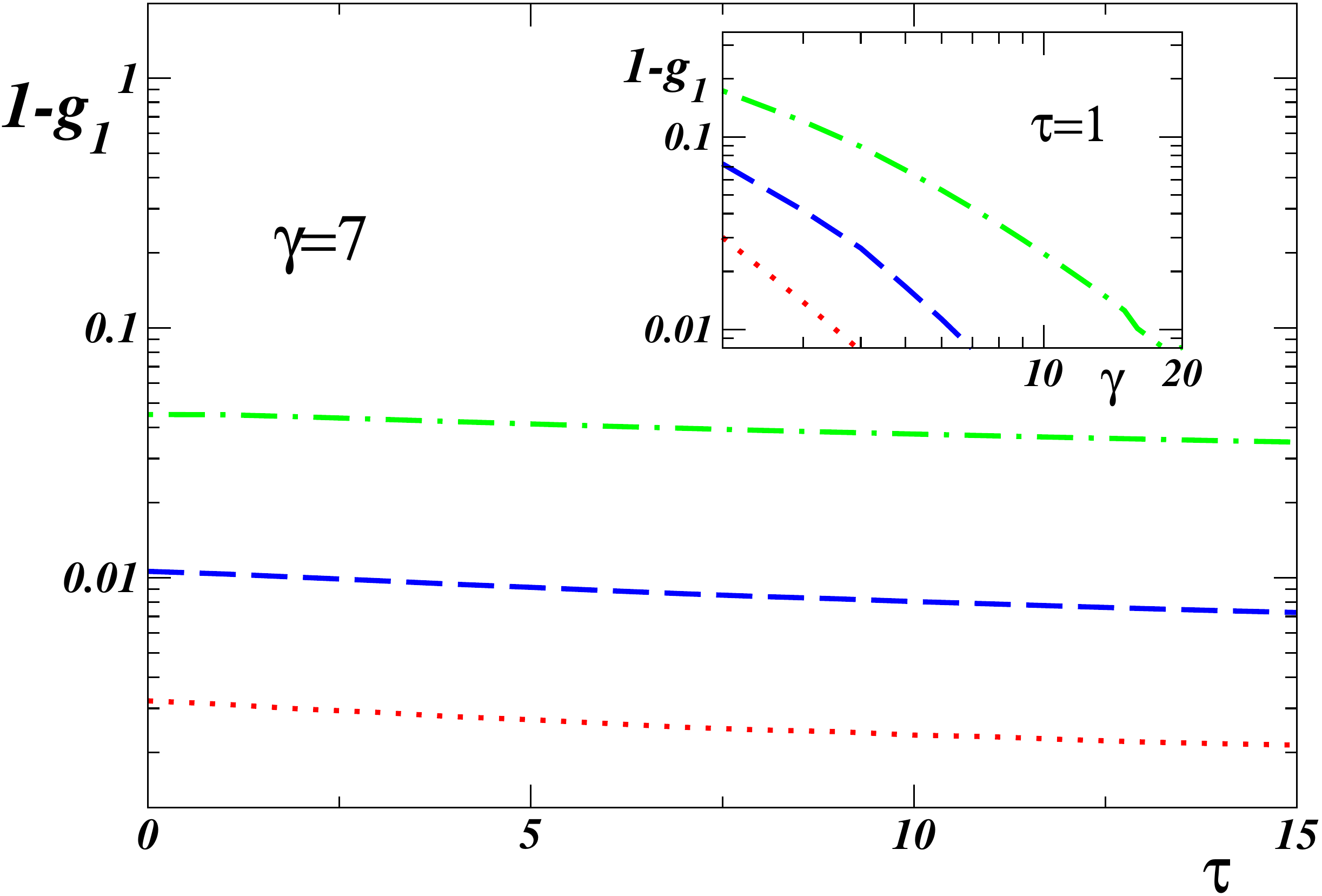}}}
\caption{Deviations $1-g_1$ from the exact result ($g_1=1$) as a function of the scaled temperature $\tau$ for a fixed value of $\gamma=7$.
Inset: $1-g_1$ vs $\gamma$ at $\tau=1$. In both figures form factors are used up to 
$n=\textcolor{green}{4}$ (green dot-dashed), $\textcolor{blue}{6}$ (blue dashed) and $\textcolor{red}{8}$ (red dotted) particles.}
\label{fig:g1T}
\end{figure}

Before discussing the numerical results, let us derive an asymptotic
formula in the regime where $1\ll\tau\ll\gamma^2$. The second
inequality implies that the $q^2$ term in the equation for
$\tilde\eps(q)$ becomes quickly quite large while the first one implies
that $\alpha$ is a large negative number. This means that, even for $q$
close to zero, the convolution term is small and 
the leading order $\tilde\eps(q)$ is given by 
\be
\tilde\eps(q) = -\alpha+\frac{q^2\gamma^2}{\tau}\,,
\ee
so we can make an expansion in the small parameter
$\exp(-\alpha+q^2\gamma^2/\tau)$
\footnote{This is reminiscent of the
small temperature expansion of the relativistic TBA equations 
where the leading term, $m\cosh(\th)/(\kb T)$, automatically 
dominates for small $T$.}. At the leading order $g(q)=1/(2\pi\alpha)$ and 
this implies
\be
\int_{-\infty}^\infty\frac{\ud q}{2\pi}e^{\alpha-\frac{q^2\gamma^2}{\tau}}=\frac1\gamma\,,
\ee
so we arrive at the $\gamma$-independent result
\be
e^\alpha=\sqrt{\frac{4\pi}\tau}\,.
\ee
We see that for $\tau\gg1$ $\alpha$ is indeed a large negative quantity.

\begin{figure}[t]
\centerline{\scalebox{0.3}{\includegraphics{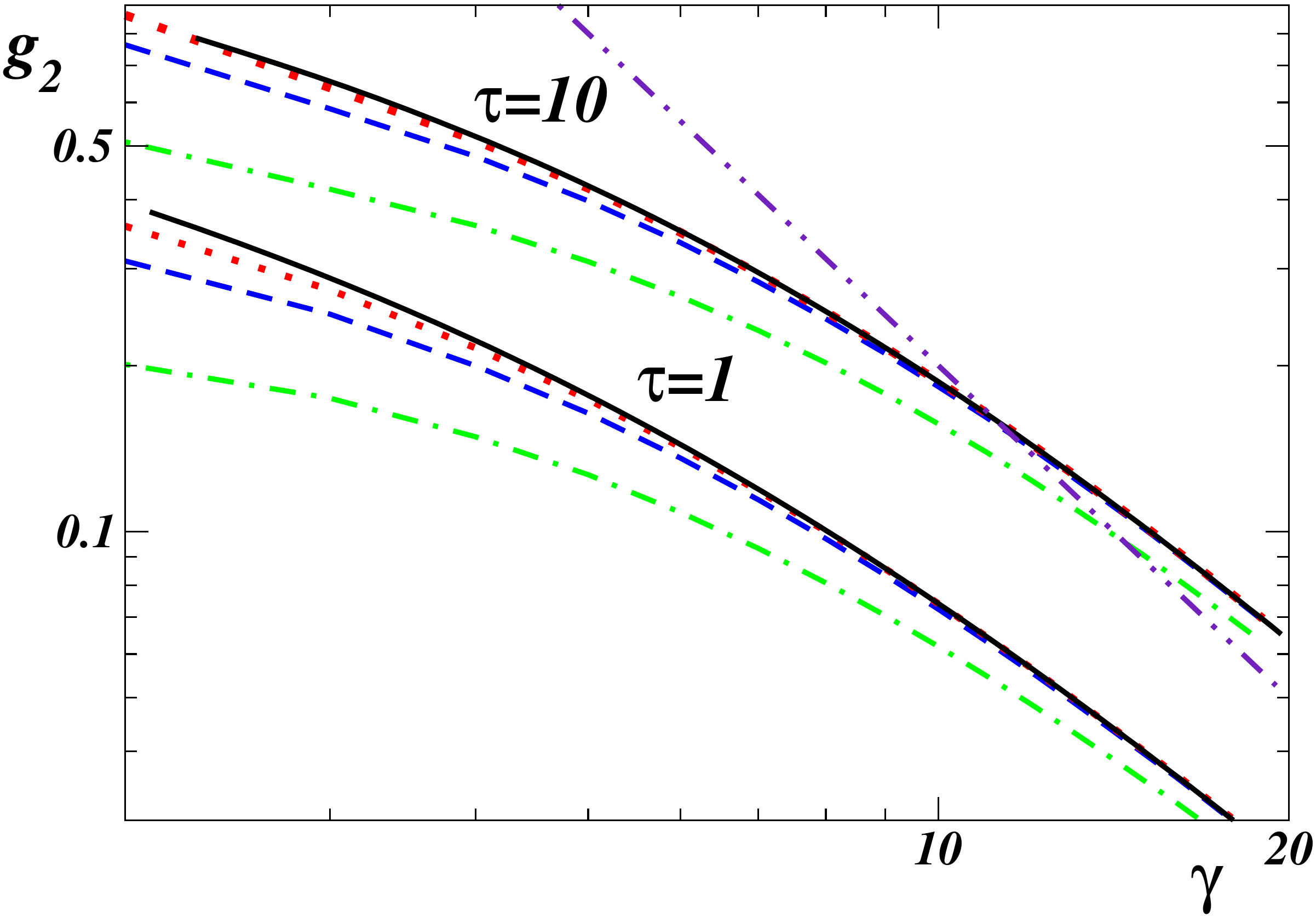}}}
\caption{$g_2$ at $\tau=1$, $10$ using form factors up to $n=\textcolor{green}{4}$, 
$\textcolor{blue}{6}$ and $\textcolor{red}{8}$ particles with 
green dot-dashed, blue dashed and red dotted lines, respectively. The
solid lines show the exact result, while the purple dot-dot-dashed line is the leading order expression \erf{eq:LO_T}.}
\label{fig:g2T}
\end{figure}

Due to the condition $\gamma\gg1$, we can again restrict ourselves to the large
$\gamma$ limit of the first non-zero term in the series
\erf{eq:formula}. Substituting the Boltzmann filling fraction
$f(\th)=e^{-\tilde\eps}$ given above, we arrive at
\be
g_k(\gamma,\tau)=\left(\frac\tau{\gamma^2}\right)^{\frac{k(k-1)}2}\,J_k\,,
\labl{eq:LO_T}
where
\be
J_k=\frac{k!}{\pi^{k/2}}
\int\ud x_1\dots\ud x_k \,e^{-\sum_{i=1}^k
  x_i^2}\,\prod_{i<j}^k(x_i-x_j)^2 = \frac{B_k}{2^{k(k-1)/2}}
\ee
with $B_{k+1}=(k+1)\Gamma(k+2)B_k$, $B_1=1$. This is exactly the
result found in \cite{gangardt}. However, as one can check
numerically, the filling fraction comes close to a Boltzmann
distribution only for such extreme parameter values as
$\gamma\sim1000$ and $\tau\sim10000$. In Figs.~\ref{fig:g2T} and
\ref{fig:g3T} one can see how large the difference is between
this leading order approximation and our result for $\tau=10$. The
improvement achieved in the determination of this quantity by the
method proposed in this paper may have an important experimental
relevance.

Let us turn now to the numerical results obtained by exactly solving
the TBA equations \erf{eq:diml}, substituting $\tilde\eps$ in the
formula \erf{eq:formula} and numerically integrating the first terms
in the series. We consider separately the computation of $g_1$, $g_2$ 
and $g_3$.

\subsubsection{Expectation value $g_1$}
\noindent
To test the reliability of our expansion at finite
temperature we computed the deviations from the exact result for the
trivial expectation value $g_1=1$. The results are plotted in
Fig.~\ref{fig:g1T}, showing that the precision does not decrease with
increasing temperature: even using only three terms of the expansion
(i.e. summing up to $n=6$ particles) the error is $\lesssim 1\%$ in
the range of temperature between $\tau=0$ and $\tau=15$. In the inset
of Fig.~\ref{fig:g1T} we plot the deviation from the exact result as a
function of the LL parameter $\gamma$ at a fixed temperature.

\subsubsection{Expectation value $g_2(\gamma,\tau)$}
\noindent
For $T>0$ the Hellmann--Feynman theorem gives 
\be
\vev{\psid\psid\psi\psi}=\frac{\ud}{\ud\lambda}\left(\frac{\tilde
  F}L\right)\,,
\labl{eq:HF2}
where the free energy can be calculated from the TBA approach \erf{eq:FE}. In
dimensionless variables
\be
g_2(\gamma,\tau)=\tau\frac{\ud}{\ud\gamma}\left(\alpha-\gamma\int_{-\infty}^\infty
\frac{\ud q}{2\pi}\,\log(1+e^{-\tilde\eps(q)})\right)\,.
\ee
We derive now a simpler expression for this. The trick is to
substitute for $1/2\pi$ under the integral the rest of Eq.~\erf{eq:gq}, then using the associativity of the convolution by an
even function and finally use the derivative of Eq.~\erf{eq:epsq}
with respect to $\gamma$. Many terms drop out and we are left with
\be
g_2=2\gamma^2\,\int_{-\infty}^\infty\ud
q\,\frac{\alpha g(q)}{1+e^{\tilde\eps(q)}}\,q^2 - \tau\,\int_{-\infty}^\infty
\frac{\ud q}{2\pi}\,\log(1+e^{-\tilde\eps(q)})\,.
\labl{eq:g2new}  
The advantage of this expression is that it is enough to solve the TBA
equations for the value of $\gamma$ we are interested in instead of evaluating
the free energy for several $\gamma$ and then differentiating it numerically. 

Our evaluations of $g_2$ at $\tau=1$ and $\tau=10$ based on the form
factor expansion are shown in Fig.~\ref{fig:g2T} together with the
exact result \erf{eq:g2new} and the leading order result
\erf{eq:LO_T}. The convergence of our series is basically as good as it
was for $T=0$ and it is clear that the asymptotic formula fails,
especially for $\gamma<10$.

\begin{figure}[t]
\centerline{\scalebox{0.3}{\includegraphics{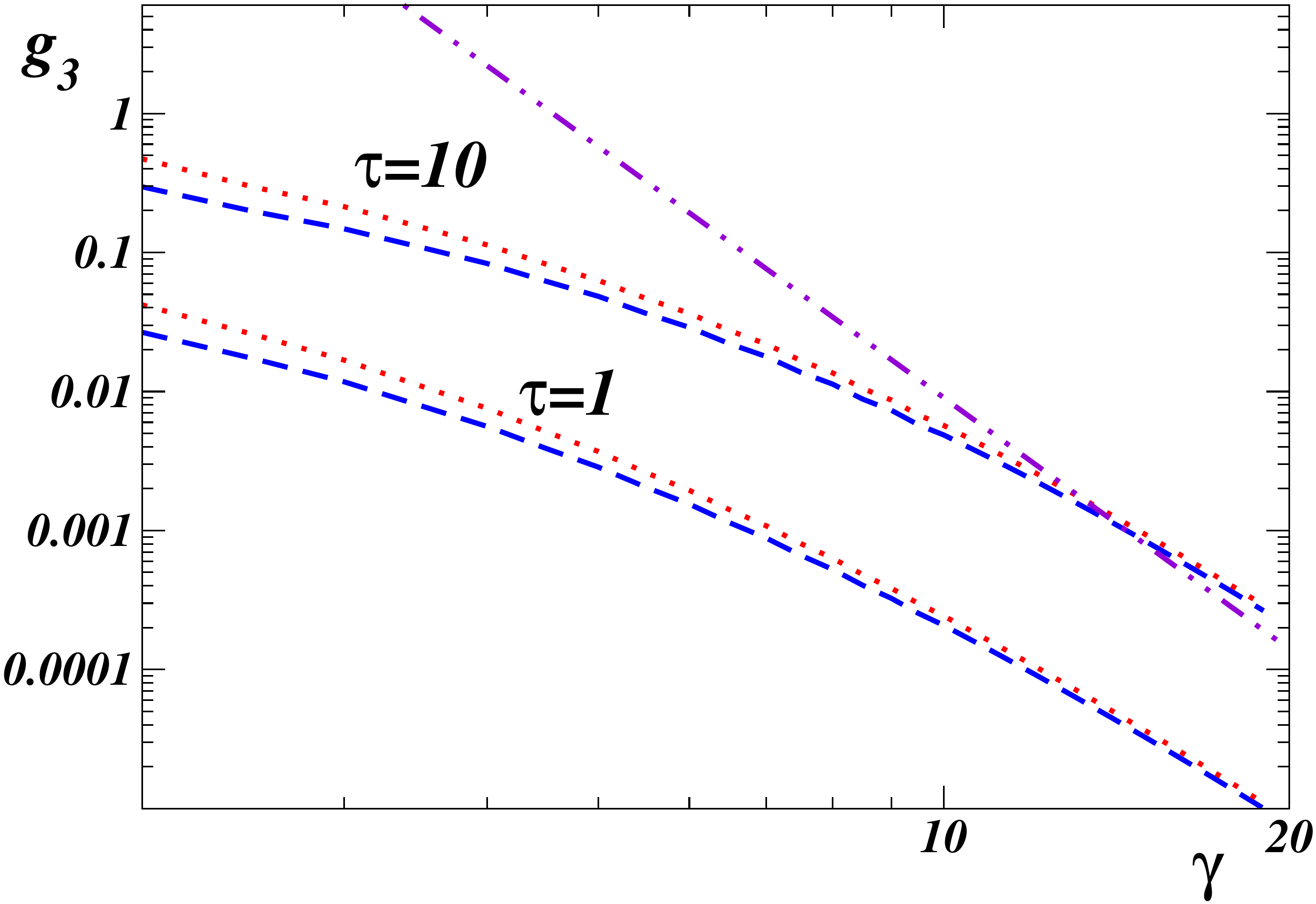}}}
\caption{$g_3$ vs $\gamma$ at $\tau= 1$ and $\tau=10$. The blue dashed and the red 
dotted lines refer to $n=\textcolor{blue}{6}$ and 
$\textcolor{red}{8}$ particles, respectively; the purple
dot-dot-dashed line shows the
asymptotic result \erf{eq:LO_T}.}
\label{fig:g3T}
\end{figure}

\subsubsection{Expectation value $g_3(\gamma,\tau)$}
\noindent
In the case of $g_2$ and of $g_3$ at $T=0$ we checked our results
using exact formulae. We learned that on the one hand our result for
$g_3$ reaches the same accuracy as $g_2$ for slightly higher values of
$\gamma$, on the other hand going to finite temperature does not spoil
the precision of our results.  Thus we are confident about the
reliability of our results for $g_3$ at $T>0$, at least for not too
small values of $\gamma$. We emphasize that neither exact results nor
approximations of precision comparable to ours exist in this case,
which renders our evaluation a new result.

\begin{figure}[t]
\centerline{\scalebox{0.3}{\includegraphics{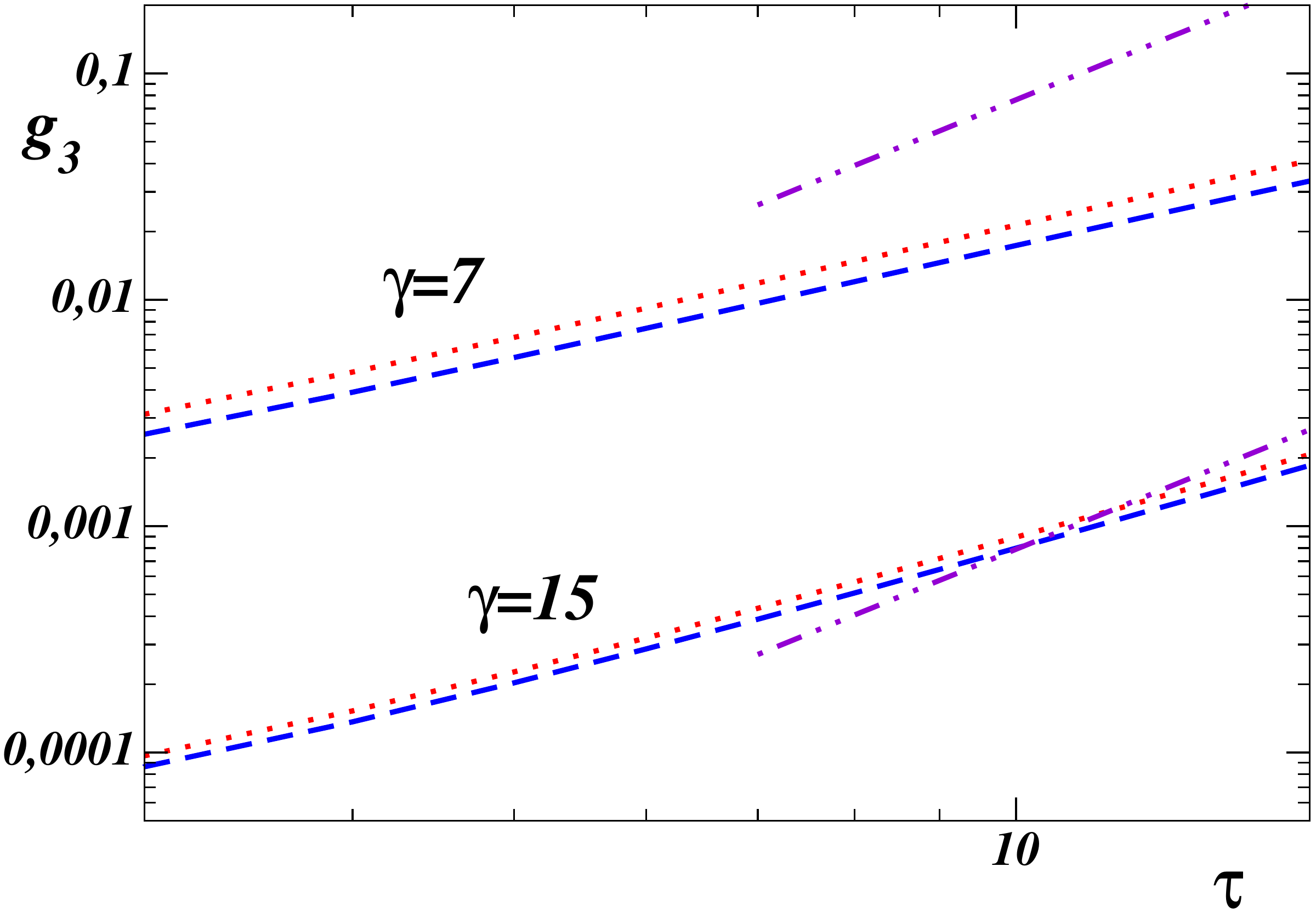}}}
\caption{$g_3$ vs the scaled temperature $\tau$ for $\gamma=7$ and 
$\gamma=15$. The blue dashed and the red 
dotted lines refer to $n=\textcolor{blue}{6}$ and 
$\textcolor{red}{8}$ particles, respectively; the purple
dot-dot-dashed lines show the
asymptotic result \erf{eq:LO_T}.}
\label{fig:g3gammafixed}
\end{figure}

Fig.~\ref{fig:g3T} shows $g_3$ as a function of $\gamma$ at fixed
temperatures $\tau=1$ and $\tau=10$. In the figure the asymptotic result
\erf{eq:LO_T}, valid for large temperature and coupling, is also
plotted: one can see that even for $\tau=10$ the result \erf{eq:LO_T}
does not give the exact asymptotic behavior (which is only reached
for very large values of the scaled temperature
$\tau$). Fig.~\ref{fig:g3gammafixed} shows instead $g_3$ as a function
of $\tau$ at fixed values of $\gamma=7$ and $\gamma=15$. The asymptotic formula
\erf{eq:LO_T} is different from our result by a factor of $\sim 10$ at
$\gamma=7$ and $\tau=10$.

\section{Conclusions}
\label{sec:concl}
\noindent 
In this paper we have shown that the Lieb--Liniger model, describing
one-dimensional interacting bosons, can be obtained as a
non-relativistic limit of an integrable relativistic field theory, the
Sinh--Gordon model. In this limit, the $S$-matrix,
the Lagrangian and the Thermodynamical Bethe Ansatz equations of the
sh-G model reduce to those of the LL model. We have also shown that the pseudo-energies of the sh-G TBA (which are actually the energies of the excitations above the vacuum) become massless modes in the non-relativistic limit, 
in agreement with the hydro-dynamical description of the LL model given by bosonization.  

The mapping between the two models proved to provide an efficient method to 
compute expectation values in the LL model by using the form
factor expansion of the expectation values of the relativistic
counterpart. The main advantage of using the form factors of the
relativistic integrable sh-G model is that, as for any relativistic quantum field theory, 
its form factors obey a set of stringent constraints that permits the determination of their exact 
expressions. Moreover, the quantum integrability of the relativistic model 
allowed us to employ the rich collection of results valid for
these systems, like the TBA and the
LeClair--Mussardo formalism.

Using these two formalisms of the relativistic sh-G model (form
factors and TBA), we computed the
expectation values of the LL model. The method works equally well at
$T=0$ and $T\neq 0$ where the series expansion presents a remarkable
convergence behavior for finite values of the LL parameter
$\gamma$. The computation of one-point correlators was presented
in detail, as well as the comparison with the known results available in
the literature. In particular, we have determined the expectation value
$g_3(\gamma,\tau)$ at finite temperature, for which there had been
only asymptotic analytic results in the literature. This quantity is
related to the recombination rate of the atomic gas and thus to
the lifetime of the experiments.

It would be interesting to analyze the possibility of extending our method to other
non-relativistic strongly correlated systems, identifying their relativistic counterparts. 
An equally important direction would be to see how the methods presented here can be used to compute space- and time-dependent two-point functions. 

\vspace{3mm}
{\it Acknowledgements:} We 
wish to thank G\'abor Tak\'acs, K\'alm\'an Szab\'o,
Bal\'azs Pozsgay, Fabian Essler and Grigory Astrakharchik 
for inspiring discussions. This work was supported by the grants
INSTANS (from ESF) and 2007JHLPEZ (from MIUR).

\appendix

\section{Determination of the coefficients $A^{k}_{l}$}
\label{app:A}
\noindent
To find the coefficients $A^{k}_{l}$ at level $k$ in Eq.~(\ref{A_k_l}) 
we can proceed in
the following way. If we already know the form factors
$F^{\no{\,\phi^{j}\,}}_{l}$ for $j<k$, we can extract the $A^{k}_{l}$'s one
by one from $\tilde F^{k}_{l}$:
\bes
\begin{align}
A^{k}_2&=\frac{\tilde F^{k}_2}{F^{\no{\,\phi^2\,}}_2}=\frac{\tilde F^{k}_2}{\tilde F^2_2}\,,\\
A^{k}_4&=\frac{\tilde F^{k}_4-A^{k}_2\,F^{\no{\,\phi^2\,}}_4}{F^{\no{\,\phi^4\,}}_4}\,,\\
&\;\;\,\vdots\nonumber\\
A^{k}_{l}&=\frac{\tilde F^{k}_{l} - \sum_{j=2,4,\dots}^{k-2}A^{k}_{j}\,F^{\no{\,\phi^{j}\,}}_{l}}{F^{\no{\,\phi^{k}\,}}_{k}}\,.
\end{align}
\esu
Once these coefficients are known, all the
form factors of $\no{\phi^{k}}$, including the ones with $n>k$
particles can be obtained from $\tilde F^{k}_{n}$. These will be
needed at higher levels. For example
\be
F^{\no{\,\phi^6\,}}_{10}=\tilde
F^6_{10}-A^6_4\,F^{\no{\,\phi^4\,}}_{10}-A^6_2\,F^{\no{\,\phi^2\,}}_{10}=\tilde
F^6_{10}-A^6_4\left(\tilde F^4_{10}-A^4_2\,\tilde
F^2_{10}\right)-A^6_2\,\tilde F^2_{10}\,.
\ee
As an example we give the rules for the ``operator
mixing'' \erf{eq:mixing} explicitly at the
first four even levels:
\bes
\begin{align}
\tilde{\phi^2}&=\no{\phi^2}\,,\\
\tilde{\phi^4}&=\no{\phi^4}-4\,\frac{\pi^2\alpha^2}{g^2}\no{\phi^2}\,,\\
\tilde{\phi^6}&=\no{\phi^6}-20\,\frac{\pi^2\alpha^2}{g^2}\no{\phi^4}+16\,\frac{\pi^4\alpha^4}{g^4}\no{\phi^2}\,,\\
\tilde{\phi^8}&=\no{\phi^8}-56\,\frac{\pi^2\alpha^2}{g^2}\no{\phi^6}+336\,\frac{\pi^4\alpha^4}{g^4}\no{\phi^4}-64\,\frac{\pi^6\alpha^6}{g^6}\no{\phi^2}\,.
\end{align}
\esu

\section{Explicit formulas for the connected form factors}
\label{sec:Fexpl}
In this Appendix we explicitly go through the steps required for the
calculation of the limit of the form factors $\tilde
F^{\no{\,\phi^{2k}\,}}_{2l}(\{p_i\})_\text{conn}$ that enter into the
formula \erf{eq:formula}, and then we list the first few of them. 

We consider explicitly only the simplest non-trivial
case, $\tilde F^{\no{\,\phi^2\,}}_{4}(p_1,p_2)_\text{conn}$. We start
from the four-particle form factor of the exponential operator
\erf{eq:FFexp}:
\begin{multline}
F_4(k)=\langle0| e^{kg\phi}|\th_1,\th_2,\th_3,\th_4\rangle =\\
[k]\left(\frac{4\sin(\pi\alpha)}{\cal N}\right)^2
\left([k]^3\sigma_1\sigma_2\sigma_3-[k-1][k][k+1](\sigma_3^2-\sigma_1^2\sigma_4)\right) \prod^4_{i<j}\frac{F_\text{min}(\th_i-\th_j)}{x_i+x_j}\,,
\end{multline}
where the elementary symmetric polynomials \erf{eq:sigma} are to be
understood as $\sigma_k=\sigma^{(4)}_k$ and we recall that
$x_i=\exp(\th_i)$ and $[k]=\sin(k\pi\alpha)/\sin(\pi\alpha)$.
The form factor of $\no{\phi^2}$ is given by the ${\mc O}(k^2)$ term
in the Taylor-expansion in $k$ (for $\no{\phi^2}$ there is no mixing \erf{eq:mixing}):
\be
F^{\no{\,\phi^2\,}}_{4}(\th_1,\th_2,\th_3,\th_4)=
-\frac{32\pi^2\alpha^2}{{\cal N}^2g^2}(\sigma_3^2-\sigma_1^2\sigma_4)\prod^4_{i<j}\frac{F_\text{min}(\th_i-\th_j)}{e^{\th_i}+e^{\th_j}}\,.
\ee
To perform the connected limit \erf{eq:conndef} we first recall that the
product of the minimal form factors in this limit is given by \erf{eq:fmincon}. For
the rest of the formula we write $x_3 = -x_2 + i \eta_2,\;x_4 = -x_1 +
i \eta_1$ and we expand the polynomials in the numerator and denominator to obtain the finite part:
\begin{multline}
{\cal F}\left[\lim_{\eta_1\to0,\eta_2\to0} 
\frac{x_2^2(x_1^2+x_2^2)\,\eta_1^2 + 4x_1^2x_2^2\,\eta_1\eta_2  + x_1^2(x_1^2+x_2^2)\,\eta_2^2+{\cal O}(\eta^3)}
{(x_1^4-2x_1^2x_2^2+x2^4)\,\eta_1\eta_2+{\cal O}(\eta^3)}\right]\\
={\cal F}\left[\frac{4x_1^2x_2^2}{(x_1^2-x_2^2)^2} +
\frac{x_2^2(x_1^2+x_2^2)}{(x_1^2-x_2^2)^2}\frac{\eta_1}{\eta_2} +
\frac{x_1^2(x_1^2+x_2^2)}{(x_1^2-x_2^2)^2}\frac{\eta_2}{\eta_1} + {\mc O}\left(\frac{\eta_1^2}{\eta_2};\frac{\eta_2^2}{\eta_1}\right)\right]=\frac{4x_1^2x_2^2}{(x_1^2-x_2^2)^2}
=\frac1{\sinh^2\th_{12}}\,.
\end{multline}
Collecting all the terms we obtain
\be
F^{\no{\,\phi^2\,}}_{4}(\th_1,\th_2)_{\text{conn}}=
\frac{32\pi^2\alpha^2}{{\cal N}^2g^2}\,\frac1{\sinh^2\th_{12}}\, \frac{{\cal N}^2\sinh^2\th_{12}}{\sinh^2\th_{12}+\sinh^2(\pi\alpha)}=
\frac{32\pi^2\alpha^2}{g^2}\,\frac1{\sinh^2\th_{12}+\sinh^2(\pi\alpha)}\,.
\ee
We note that another way to arrive at this result is to calculate
first the connected form factors of $\exp(kg\phi)$ and then to extract
the ${\mc O}(k^2)$ term to obtain the connected form factors
of $\no{\phi^2}$.

Finally, after substituting $\th_i\to p_i/mc$ and using the definition of $\alpha$ \erf{eq:alpha},
we can perform the double limit (\ref{eq:limit},\ref{eq:id}). We list below
the result together with the first few non-relativistic form factors obtained in this way.
We use the notation $p_{ij}=p_i-p_j$, while $\sum_P$ denotes a sum over permutations of $\{i,j,k\}=\{1,2,3\}$.
\bes
\begin{align}
\tilde F^{\no{\,\phi^{2}\,}}_{2}\,\!\!_\text{conn}&=\frac1\hbar\,,\\
\tilde F^{\no{\,\phi^{2}\,}}_{4}(p_1,p_2)_\text{conn}&=\frac{8m\lambda}{4m^2\lambda^2+\hbar^2p_{12}^2}=\frac2\hbar\tilde\fii(p_{12})\,,\\
\tilde F^{\no{\,\phi^{2}\,}}_{6}(p_1,p_2,p_3)_\text{conn}&=\frac{32 \hbar m^2 \lambda^2 (12 m^2 \lambda^2 + 
   \hbar^2 (p_{12}^2+p_{13}^2+p_{23}^2))}
{(4 m^2 \lambda^2 + 
   \hbar^2 p_{12}^2) (4 m^2 \lambda^2 + 
   \hbar^2 p_{13}^2) (4 m^2 \lambda^2 + \hbar^2
   p_{23}^2)}=\frac1\hbar \sum_P \tilde\fii(p_{ij})\tilde \fii(p_{jk})\,.
\end{align}
\esu
\bes
\begin{align}
\tilde F^{\no{\,\phi^{4}\,}}_{4}(p_1,p_2)_\text{conn}&=\frac1{m\lambda\hbar}\,\tilde\fii(p_{12})p_{12}^2\,,\\
\tilde
F^{\no{\,\phi^{4}\,}}_{6}(p_1,p_2,p_3)_\text{conn}&=8\hbar m\lambda\,\frac{(p_{12}^2+p_{13}^2+p_{23}^2)(8m^2\lambda^2+\hbar^2(p_{12}^2+p_{13}^2+p_{23}^2))}{(4 m^2 \lambda^2 + 
   \hbar^2 p_{12}^2) (4 m^2 \lambda^2 + 
   \hbar^2 p_{13}^2) (4 m^2 \lambda^2 + \hbar^2
   p_{23}^2)}= \nonumber\\
&=\frac1{2m\lambda\hbar} \sum_P \tilde\fii(p_{ij})\tilde\fii(p_{jk})p_{ik}^2\,.
\end{align}
\esu
\begin{align}
\tilde F^{\no{\,\phi^{6}\,}}_{6}(p_1,p_2)_\text{conn}&=36\hbar^3\frac{p_{12}^2\,
p_{23}^2 \,p_{13}^2}{(4 m^2 \lambda^2 + 
   \hbar^2 p_{12}^2) (4 m^2 \lambda^2 + 
   \hbar^2 p_{23}^2) (4 m^2 \lambda^2 + \hbar^2
   p_{13}^2)}=\nonumber\\
&=\frac9{16m^3\lambda^3}\tilde\fii(p_{12})\tilde\fii(p_{23})\tilde\fii(p_{13})p_{12}^2\,p_{23}^2 \,p_{13}^2\,.
\end{align}

\end{document}